%% file: main_arxiv.tex
\documentclass{article} 
\usepackage{iclr2027_conference,times}

\input{math_commands.tex}

\usepackage{hyperref}
\usepackage{url}

\usepackage{algorithm}
\usepackage{enumitem}
\usepackage{newfloat}
\usepackage{listings}
\usepackage[most]{tcolorbox}
\usepackage{xcolor}  
\usepackage{booktabs}
\usepackage{multirow}
\usepackage{amsmath, amssymb}
\usepackage{booktabs}
\usepackage[table]{xcolor}
\usepackage{tabularx}
\usepackage{longtable}
\usepackage{array} 
\usepackage{pifont}
\usepackage{algorithm}
\usepackage{algpseudocode} 

\usepackage{wrapfig}
\usepackage{graphicx} 
\usepackage{xspace}
\usepackage{enumitem}
\usepackage{booktabs}         
\usepackage{multirow}         
\usepackage{colortbl}         
\usepackage[table]{xcolor}    
\usepackage{graphicx}         
\usepackage{algorithm} 
\usepackage{algpseudocode} 
\usepackage{amsmath, amsfonts, amssymb} 
\usepackage{amsthm}
\usepackage[most]{tcolorbox}
\usepackage{enumitem}
\usepackage{array}

\newtcolorbox{promptbox}[1]{
  colback=black!3,
  colframe=black!60,
  title=#1,
  breakable,
  left=2mm,
  right=2mm,
  top=1mm,
  bottom=1mm,
  before skip=6pt,
  after skip=8pt
}

\usepackage{booktabs} 
\usepackage{adjustbox} 
\usepackage{times}
\usepackage{latexsym}
\usepackage{colortbl}
\usepackage{xcolor}
\usepackage{threeparttable}
\usepackage{tikz}
\usepackage{subcaption}  
\usepackage{threeparttable}
\usepackage{hyperref}
\theoremstyle{remark}

\definecolor{citecolor}{RGB}{2, 56, 189}
\definecolor{myred}{RGB}{207,62,62}
\definecolor{mygreen}{RGB}{112,173,71}
\definecolor{mynewred}{RGB}{178, 56, 40}
\definecolor{rebuttal_blue}{HTML}{1685a9}
\definecolor{rebuttal_purple}{HTML}{9932cd}
\definecolor{rebuttal_red}{HTML}{ef7a82}
\definecolor{avgblue}{RGB}{210,230,250}

\usepackage[dvipsnames]{xcolor}

\newcommand{\M}{\textsc{GitHarness}}
 \newcommand{\dataset}{\textsc{MTAgentBench}}

\usepackage{booktabs}
\usepackage[table]{xcolor}
\usepackage{colortbl}
\usepackage{multirow}

\usepackage{cleveref}

\usepackage{xcolor}

\usepackage{pifont}

\title{\textsc{GitHarness}: Git Init Your Harness Working Memory for Perpetual User Requirements}

\author{
\textbf{Zhibang Yang\textsuperscript{1,2,3}\thanks{Equal contribution.}},
\textbf{Xinke Jiang\textsuperscript{1,2,3}\footnotemark[1]},
\textbf{Yuxuan Liu\textsuperscript{1}\footnotemark[1]},
\textbf{Mingyu Zhang\textsuperscript{1}},\\
\textbf{Zhixin Zhang\textsuperscript{1,2,3}},
\textbf{Zhengxing Song\textsuperscript{1}},
\textbf{Yue Fang\textsuperscript{1,2,3}},
\textbf{Guohong Qiu\textsuperscript{1}},\\
\textbf{Ruiqing Li\textsuperscript{1}},
\textbf{Xu Chu\textsuperscript{2,3,4}\thanks{Corresponding authors.}},
\textbf{Junfeng Zhao\textsuperscript{2,3}\footnotemark[2]},
\textbf{Yasha Wang\textsuperscript{1,5}\footnotemark[2]}\\
\textsuperscript{1}National Engineering Research Center of Software Engineering, Peking University, Beijing, China\\
\textsuperscript{2}School of Computer Science, Peking University, Beijing, China\\
\textsuperscript{3}Key Laboratory of High Confidence Software Technologies, Ministry of Education, Beijing, China\\
\textsuperscript{4}Center on Frontiers of Computing Studies, Peking University, Beijing, China\\
\textsuperscript{5}Peking University Information Technology Institute (Tianjin Binhai), Tianjin, China\\
\small\texttt{\{yangzb,xinkejiang\}@stu.pku.edu.cn}}
\iclrfinalcopy

\begin{document}

\maketitle
\fancyhead{}

\begin{abstract}
LLM-based agents increasingly collaborate with users on long-horizon tasks, accumulating evidence, code, and drafts through extensive search, reasoning, and execution. As users inspect these results, they may supply missing information (\emph{\textbf{requirement completion}}), introduce new requirements (\emph{\textbf{requirement elicitation}}), or revise existing ones (\emph{\textbf{requirement shift}}). These changes often affect only part of the accumulated work, yet agents may carry forward obsolete information or turn local revisions into global rewrites. Existing approaches clarify current intent without determining how prior work should change, or reuse execution histories under a fixed objective. We address this gap by formulating perpetual-requirement collaboration as joint requirement tracking and local update. We introduce \textsc{\textbf{GitHarness}}, a pluggable \underline{\textbf{Git}}-style framework that organizes requirement states and their corresponding \textbf{\underline{harness}} work states into a branchable version history. A trainable Git Agent resolves requirement changes and selects a semantically compatible historical state. A unified version interface then restores that state and creates a new branch, enabling the underlying harness to exclude obsolete information, inherit compatible work, and focus execution on affected parts. The Git Agent is trained through interface-level black-box reinforcement learning, with downstream harnesses and task-execution models kept fixed. We also construct \dataset{}, a verifier-preserving benchmark covering mathematical reasoning, text-to-SQL, agentic search, software engineering, and research synthesis. Experiments demonstrate strong task performance alongside effective requirement tracking, preservation of valid work, and efficient execution.
\end{abstract}

\input{1_Introduction}

\input{2_Relatedwork}


\input{4_Method}

\input{5_Experiment}

\input{6_conclusion}

\subsection*{AI use statement}

Large language models were used for language polishing, literature discovery, synthetic trajectory construction, figure-design assistance, and AI-assisted programming. LLMs helped transform public benchmark tasks into multi-turn requirement trajectories and provided visual drafts, code suggestions, and debugging support. All generated data, figures, citations, and code were reviewed and verified by the authors, with benchmark verifiers and human inspection used to validate synthetic trajectories. The research questions, methodology, experimental design, analyses, and conclusions were determined by the authors.

\subsection*{Ethics statement}

All experiments use publicly available benchmarks, including GSM8K, BIRD, BrowseComp-Plus, SWE-bench Verified, and DeepResearch Bench, under their respective licenses and terms. The study involves no human or animal subjects and uses no private conversations or personally identifiable information. The multi-turn interactions are generated from public benchmark tasks using author-designed construction protocols and LLM assistance, rather than collected from real users. Software-engineering tasks are executed in isolated benchmark environments to limit unintended effects on external systems.

\subsection*{Reproducibility statement}

We will release the implementation of \M, the constructed multi-turn trajectories, and the metadata required to reproduce their requirement transitions and verification. The appendix reports, prompts, model configurations, training settings, baseline implementations, and evaluation procedures. Evaluation scripts, LLM-judge configurations, and commands for reproducing the main results will be added to the anonymized repository at \url{https://anonymous.4open.science/r/GitHarness-7A2E/}. All experiments rely on publicly accessible benchmarks and model interfaces; no private data is required.

\bibliography{iclr2027_conference}
\bibliographystyle{iclr2027_conference}

\appendix

\input{7_appendix}

\end{document}

%% file: math_commands.tex
\usepackage{amsmath,amsfonts,bm}

\def\eqref#1{equation~\ref{#1}}

\def\1{\bm{1}}

\DeclareMathAlphabet{\mathsfit}{\encodingdefault}{\sfdefault}{m}{sl}
\SetMathAlphabet{\mathsfit}{bold}{\encodingdefault}{\sfdefault}{bx}{n}



%% file: 1_Introduction.tex
\section{Introduction}
\label{sec:introduction}

LLM-based agents \citep{bubeck2023paper} can now carry out long-horizon tasks such as software engineering, open-world search, and deep research \citep{guo2025deepseek,jaech2024openai,jimenez2024swe,zhou2024webarena,li2025webweaver,du2025deepresearch}. During execution, they gather evidence, develop plans, invoke tools, and produce intermediate results such as code, queries, and report drafts. Yet users rarely provide a complete and immutable task specification at the outset. As they inspect intermediate results, they may supply missing task information (\emph{\textbf{requirement completion}}), introduce new requirements inspired by the results (\emph{\textbf{requirement elicitation}}), or revise, replace, or withdraw existing requirements (\emph{\textbf{requirement shift}}) \citep{tack2026llms,subramonyam2024bridging,lee2022evaluating,faltings2023interactive}.

\begin{figure*}
    \centering
    \includegraphics[width=0.9\textwidth]{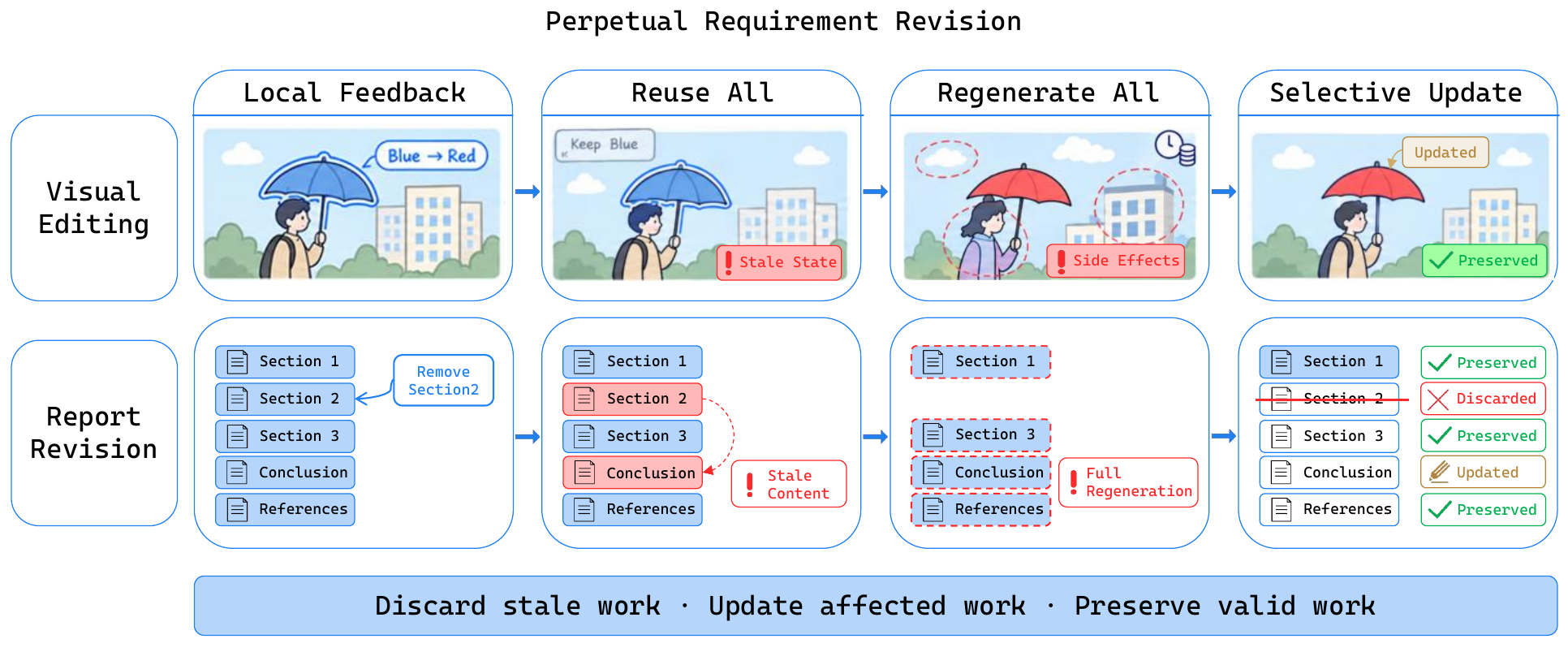}
    \caption{When user feedback changes the umbrella color or removes a report section, reusing all prior work leaves stale content, while regenerating everything may alter unaffected parts. Selective updating limits changes to the affected work.}
    \label{fig:challenges}
\end{figure*}

Unlike conventional multi-turn chat \citep{lee2025multiversemultiturnconversationbenchmark}, the context of long-horizon agent collaboration includes not just dialogue but also the results of \textbf{extensive search, reasoning, and execution}. Subsequent user feedback often targets only part of this accumulated work. An agent must \textbf{track active requirements and reuse valid context and results}. In practice, however, a request to fix a single bug may trigger changes to unrelated files across the workspace~\citep{zhu2026modelseditmuchfidelity}, while a request to refine one visual detail may alter the entire image (Figure~\ref{fig:challenges}). A local revision can thus become a global rewrite, unnecessarily changing work that was already correct. This motivates our central research question: \emph{\textbf{How can an agent harness adapt to changing user requirements while excluding obsolete information, preserving unaffected work, and updating only affected parts?}}

Retaining the full history risks carrying forward obsolete requirements and results, whereas restarting from current requirements discards potentially reusable work. Existing methods address this trade-off only partially. \ding{182} Context retrieval, summarization, and intent reconstruction help clarify the current request \citep{dongre2025drift,liu2026intentmismatchcausesllms,su2026u}, but do not directly determine how prior work should be reused, potentially requiring re-execution at substantial token and time cost. \ding{183} Trajectory organization, compression, and restoration support the reuse of intermediate work \citep{hu2025hiagent,li2026sculptor,hu2026sam,zhuang2026agentrewind}, but typically assume a fixed task objective; when requirements change, superseded requirements and their associated results may continue to influence execution \citep{laban2026llms,tack2026llms,li2025structflowbench}. The missing link is an explicit connection between requirement changes and decisions about which prior work to preserve, discard, or recompute.

Addressing this gap requires two linked decisions: \emph{\textbf{where to resume}}---identifying the currently effective requirements and selecting a compatible historical state; and \emph{\textbf{what to update}}---determining which parts of the prior work must change and which can be preserved. \textbf{Version control} offers a natural abstraction for these decisions. Git preserves versions through commits, supports branching from past states, and records local changes as diffs. This suggests a way to \emph{git init} an agent's working memory: organize requirements and their corresponding work as versioned states, and treat each new piece of feedback as a branch update from an appropriate historical state.

To this end, we introduce \textsc{GitHarness}, a pluggable Git-style version-tracking framework that supports dynamic user requirements across diverse agent harnesses. It couples each effective requirement state with its corresponding work state in a branchable version history. A trainable Git Agent identifies effective requirements, interprets their changes, and locates a semantically compatible historical state, translating multi-turn feedback into explicit version transitions. \textsc{GitHarness} then restores the selected work state and creates a new branch, allowing the underlying harness to exclude obsolete states, inherit compatible work, and focus further execution on the affected parts. A unified version interface maps these operations onto the native work states of different harnesses without redesigning their working memory or execution procedures. We train the Git Agent through interface-level black-box reinforcement learning, using the correctness of requirement resolution, change interpretation, and historical-state localization as feedback while keeping the downstream harnesses and task-execution models fixed. Our contributions are threefold:
\begin{itemize}[leftmargin=*,noitemsep,topsep=2pt]
    \item We formulate long-horizon agent collaboration under evolving requirements as a joint problem of requirement tracking and selective work reuse. The formulation covers requirement completion, elicitation, and shift, and connects changes in user requirements to decisions about where to resume execution and what to update.

    \item We propose \textsc{GitHarness}, a pluggable framework that couples requirement states and work states in a branchable version history. A trainable Git Agent interprets requirement changes and selects compatible historical states, while a unified version interface supports restoration and continued execution across different harnesses without retraining the underlying task executors.

    \item We construct \dataset{}, covering mathematical reasoning, text-to-SQL, agentic search, software engineering, and research synthesis under diverse requirement-change trajectories. The dataset provides structured annotations of requirement states, version transitions, and task completion, together with an evaluation protocol for cross-task and compositional requirement changes.
\end{itemize}

%% file: 2_Relatedwork.tex
\section{Related Work}
\label{sec:related_work}

\paragraph{Dynamic user requirement tracking.}
A user request may be disclosed, extended, or revised over several turns, making the latest utterance insufficient on its own. Existing methods maintain user-side history by appending previous turns, retrieving relevant dialogue \citep{zhang2025dhragdynamichistoricalcontextpowered}, summarizing context, rewriting conversational queries \citep{yuan2025query}, or modeling relations such as follow-up, correction, expansion, and recall \citep{li2025structflowbench}. Intent-mismatch methods further separate intent mediation from task execution \citep{liu2026intentmismatchcausesllms}, while U-Fold constructs an intent-aware dialogue summary and a task-relevant tool log for evolving user intent \citep{su2026u}. Despite these advances, models remain anchored to early assumptions under progressive disclosure and struggle with intent revelation, revision, and redirection \citep{laban2026llms,tack2026llms}. These approaches clarify current intent but do \textbf{not determine how changes propagate to evidence, code, drafts, and other work produced under earlier requirements}.

\paragraph{Long-horizon agent work management.}
Long-horizon agents accumulate extensive traces and intermediate artifacts during planning and tool use. Memory methods organize, compress, or retrieve this execution-side history through stack operations, subgoal-based trajectories, active context editing, or state-adaptive recall \citep{jiang2024tc,hu2025hiagent,li2026sculptor,hu2026sam}. Recovery methods diagnose consequential failures or restore aligned agent and environment checkpoints \citep{qi2026trajdebug,ma2026dover,zhuang2026agentrewind}. Related systems further introduce version-control abstractions: GCC manages context through commit, branch, merge, and retrieval \citep{wu2025git}, while GitOfThoughts stores reasoning trees as Git repositories \citep{shekar2026gitofthoughts}. These methods make prior execution easier to reuse, but select history according to the agent's current reasoning state, an execution failure, or exploration needs under an otherwise given task. They therefore \textbf{lack a requirement-driven mechanism for deciding which historical work state remains compatible when user feedback changes the task itself}. In contrast, \textsc{GitHarness} uses requirement transitions to locate compatible historical states and decide what work to preserve, discard, or recompute.

%% file: 4_Method.tex
\section{Problem Formulation}
\label{sec:problem_formulation}

We study multi-turn tasks with evolving user requirements.
At each turn~$t$, the system maintains a \emph{requirement state}~$\boldsymbol{I}_{i,t}$ (current goal and constraints) and a \emph{work state}~$\boldsymbol{W}_{i,t}$ (execution context and accumulated artifacts).
New feedback may change the goal and invalidate prior work.
The central difficulty is that the agent cannot simply re-execute from scratch at each turn; it must identify what changed, retain valid prior work, and update only affected parts.

\paragraph{Requirement changes.}
User feedback typically modifies an existing requirement rather than defining a task from scratch.
We identify three directions of change, each formalized as an edit~$\Delta\boldsymbol{I}_{i,t}$ relative to a historical anchor state:
\begin{itemize}[leftmargin=*,noitemsep,topsep=2pt]
    \item \textbf{Completion}: adds missing details consistent with the current direction.
    \item \textbf{Elicitation}: introduces new task dimensions while preserving existing requirements.
    \item \textbf{Shift}: revises or withdraws prior requirements, invalidating part of the existing state.
\end{itemize}
The anchor need not be the most recent state---users may revisit earlier decisions or extend a historical branch.
We instantiate nine fine-grained operators for data construction and evaluation; the method itself does not require change-type identification.

\paragraph{Task definition.}
Given the accepted history $\mathcal H_{i,t-1}$ and new feedback~$q_{i,t}$, the system must (1)~recover the full current requirement, (2)~select a compatible historical base, and (3)~update the work state locally:
\begin{equation}
(\widehat{\boldsymbol{Q}}_{i,t},\;\widehat b_{i,t},\;h_{i,t})
\;=\;
f\!\left(q_{i,t},\;\mathcal H_{i,t-1}\right),
\qquad
\widehat{\boldsymbol{W}}_{i,t}
\;=\;
F_{d_i}\!\left(\boldsymbol{W}_{i,\widehat b_{i,t}},\;\widehat{\boldsymbol{Q}}_{i,t},\;h_{i,t}\right).
\label{eq:task_definition}
\end{equation}
Here $\widehat{\boldsymbol{Q}}_{i,t}$ is the resolved requirement, $\widehat b_{i,t}$ the selected base version, and $h_{i,t}$ an update hint for downstream execution.
Rather than predicting which specific artifacts are affected, the system achieves locality through base selection and branch-local execution: resuming from a compatible prior state naturally confines edits to what the new requirement demands.
For evaluation, each instance carries a final-task verifier and optional turn-level verifiers:
\begin{equation}
s_i^{\mathbf{final}}
\;=\;
V^{\mathbf{base}}_{d_i}\!\left(\widehat{\boldsymbol{Y}}_{i,T_i},\;\boldsymbol{I}_{i,T_i}\right),
\qquad\;
s_{i,t}^{\mathbf{step}}
\;=\;
\widetilde V_{i,t}\!\left(\widehat{\boldsymbol{Y}}_{i,t},\;\boldsymbol{I}_{i,t},\;\Delta\boldsymbol{I}_{i,t}\right).
\label{eq:evaluation}
\end{equation}
We additionally report execution cost.

\section{Method}
\label{sec:method}

\begin{figure*}[t]
    \centering
    \includegraphics[width=0.95\textwidth]{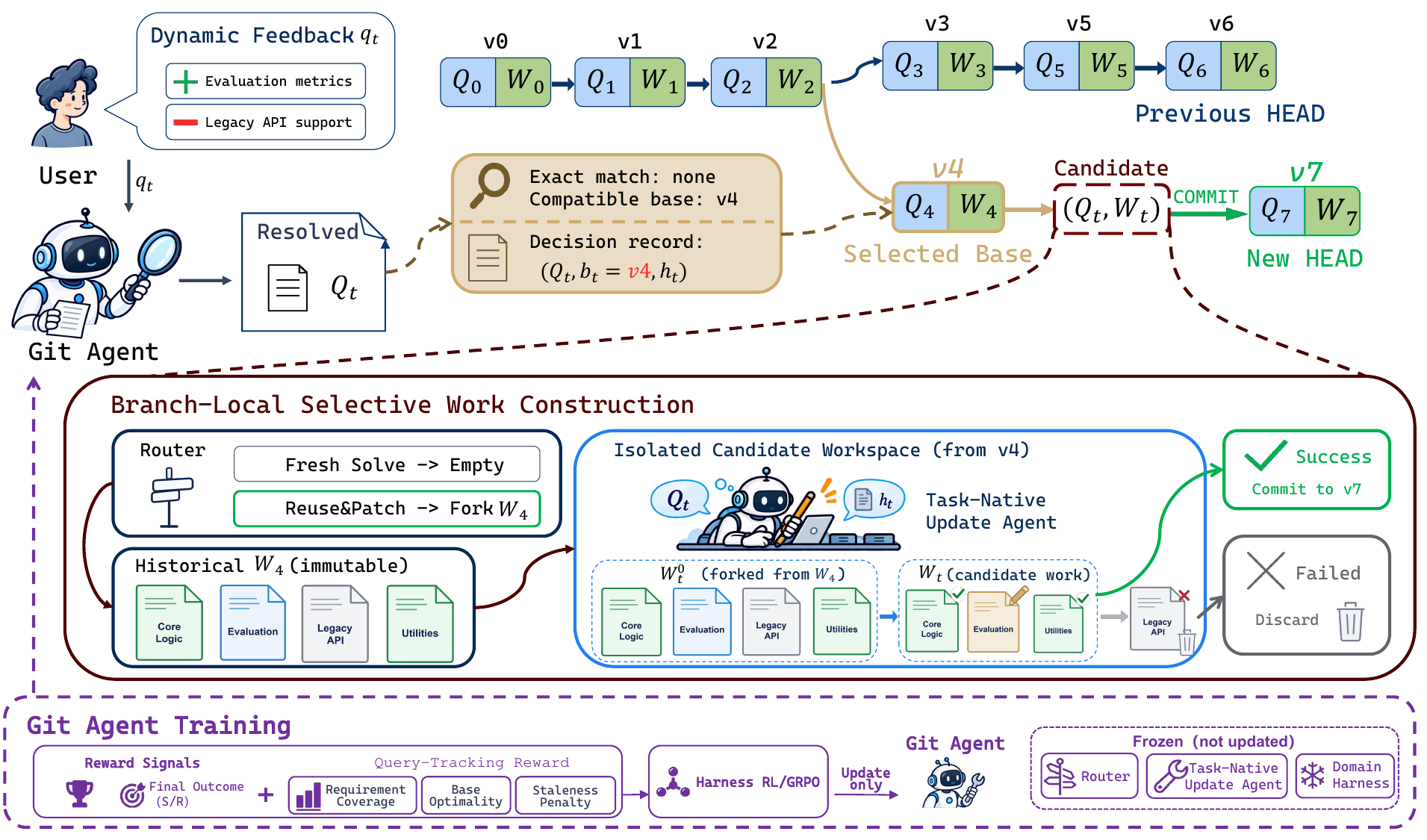}
    \caption{Framework of \M.}
    \label{fig:framework}
\end{figure*}
Flattening a growing interaction history into a single context obscures which prior states are reusable and makes selective rollback impossible.
\textsc{GitHarness} addresses this by saving a \emph{version} after each successful turn---a snapshot pairing the requirement with a recoverable checkpoint of the work state.
Versions form a directed graph~$\mathcal G_{<t}$; any past version can be restored on demand via its checkpoint~$\mathcal B(v)$.
This structure supports continuation from the current head, rollback to an earlier version, and branching when requirements diverge.
When feedback~$q_t$ arrives, \textsc{GitHarness} processes via:
\begin{enumerate}[leftmargin=*,noitemsep,topsep=2pt]
    \item \textbf{Requirement resolution} (\S\ref{sec:versioned_requirement_resolution}): A \emph{Git Agent} determines the full current requirement~$\boldsymbol{Q}_t$ and the best base version~$b_t$.
    \item \textbf{Branch execution} (\S\ref{sec:adaptive_branch_execution}): The system restores the selected base and an \emph{update agent} carries out targeted modifications.
\end{enumerate}
Separating these two concerns allows the Git Agent to focus purely on requirement understanding, while the update agent handles domain-specific execution.
Formally:
\begin{equation}
(\boldsymbol{Q}_t,\; b_t,\; h_t)
\;=\;
\pi_\theta^G\!\left(q_t,\;\mathcal G_{<t}\right),
\qquad
\boldsymbol{W}_t
\;=\;
F\!\left(\boldsymbol{W}_{b_t},\; \boldsymbol{Q}_t,\; h_t\right).
\label{eq:two_stage}
\end{equation}
Only normally terminated results are committed as new versions; failed executions leave the graph unchanged.

\subsection{Requirement Resolution}
\label{sec:versioned_requirement_resolution}

The Git Agent must jointly determine the full requirement and the best historical base---two interdependent decisions that require reasoning over the version graph.
A single-step classifier is insufficient: the agent may need to review what earlier versions achieved before deciding where to resume.
It therefore runs a short ReAct loop with three actions:
\begin{itemize}[leftmargin=*,noitemsep,topsep=2pt]
    \item \textsc{Inspect}: read a past version's requirement, gathering context without making a decision.
    \item \textsc{Commit}: declare the full current requirement and select a base version for runtime resolution.
    \item \textsc{Reuse}: select a version that already satisfies the requirement, skipping further execution.
\end{itemize}
At each reasoning step~$k$:
\begin{equation}
\begin{aligned}
a_k &\;\sim\; \pi_\theta^G\!\left(\cdot \mid q_t,\;\mathcal G_{<t},\;\eta_{<k}\right),
\qquad a_k \in \big\{\textbf{\textsc{Inspect}},\;\textbf{\textsc{Commit}},\;\textbf{\textsc{Reuse}}\big\},\\[3pt]
o_k &\;=\; \operatorname{Env}\!\left(\mathcal G_{<t},\; a_k\right).
\end{aligned}
\label{eq:react_step}
\end{equation}
A successful \textsc{Commit} or \textsc{Reuse} terminates the loop and returns the handoff $\xi_t=(\boldsymbol Q_t,b_t,v_t,h_t)$, comprising the resolved requirement~$\boldsymbol Q_t$, selected base~$b_t$, target version~$v_t$, and non-authoritative update hint~$h_t$.
\textsc{Commit} extends the main path when $b_t$ is the current head and creates a branch otherwise---branching is thus a natural by-product of base selection, not a separate action.

\subsection{Branch Execution}
\label{sec:adaptive_branch_execution}

The Git Agent has determined \emph{what} to achieve and \emph{where} to start; the remaining question is \emph{how much work} is actually needed.
Three execution modes cover the full spectrum:
\begin{equation}
z_t \;=\;
\begin{cases}
\textbf{\textsc{ExactReuse}},
    & \mathcal B(v_t)\neq\bot,\\[5pt]
\textbf{\textsc{FreshSolve}},
    & b_t=\varnothing\;\text{or}\;\mathcal B(b_t)=\bot,\\[5pt]
R_\psi(\boldsymbol{Q}_t,\;\boldsymbol{W}_{b_t}),
    & \text{otherwise}.
\end{cases}
\label{eq:exec_mode}
\end{equation}
\textsc{ExactReuse} applies when a past version already satisfies the requirement---its checkpoint is returned directly with no execution.
\textsc{FreshSolve} starts from an empty workspace when no usable base exists.
When a partially compatible base is available, the router~$R_\psi$ inspects a lightweight workspace manifest and chooses between \textsc{FreshSolve} and \textsc{ReuseAndPatch}; the latter forks the base checkpoint and applies targeted modifications, preserving unaffected work.

For the two modes that require execution, the update agent initializes and then updates a candidate state:
\begin{equation}
\boldsymbol{W}_t^{\,0}
\;=\;
\begin{cases}
\operatorname{Empty}(),
    & z_t = \textbf{\textsc{FreshSolve}},\\[4pt]
\operatorname{Fork}(\boldsymbol{W}_{b_t}),
    & z_t = \textbf{\textsc{ReuseAndPatch}},
\end{cases}
\qquad
\widetilde{\boldsymbol{W}}_t
\;=\;
\operatorname{Execute}\!\left(\boldsymbol{W}_t^{\,0},\; \boldsymbol{Q}_t,\; h_t\right).
\label{eq:execute}
\end{equation}
On normal termination the candidate is committed as a new version in the graph; on failure, the graph remains unchanged. Historical checkpoints are immutable, ensuring that failed executions never corrupt prior states.


\subsection{Training with Harness RL}
\label{sec:harness_rl_training}

Downstream quality depends on the Git Agent's requirement and base selection: even a perfect update agent cannot recover from errors in either.
We therefore train only~$\pi_\theta^G$ via reinforcement learning, treating the update agent and domain harness as a fixed environment.
We adopt Harness RL~\citep{jiang2026harness} to extract trainable segments from structured rollouts. For rollout capture, a single rollout involves interleaved calls to the Git Agent and the update agent.
Every Git Agent call is recorded as an interface call record~$\xi_n = (X_n, Y_n, \boldsymbol\ell_n^{\mathbf{roll}})$---tokenized input, sampled output, and per-token rollout log-probabilities.
Between calls, the harness dispatches execution to the update agent and assembles resulting observations into the Git Agent's next context.
Only Git Agent calls are included in the training mask; all other calls contribute context but receive no gradient.

\paragraph{Reward design.}
We combine a sparse outcome reward with a dense query-tracking reward, both attributed exclusively to Git Agent decisions.
The outcome reward checks final-artifact correctness:
\begin{equation}
R^{\mathbf{out}}_e
\;=\;
V^{\mathbf{base}}_{d_i}\!\left(\widehat{\boldsymbol{Y}}_{i,T_i},\; \boldsymbol{I}_{i,T_i}\right).
\label{eq:outcome_reward}
\end{equation}
Because this signal cannot distinguish which turns contributed to success or failure, we introduce a turn-level query-tracking reward.
At each turn~$t$, an LLM-based judger~$J_\phi$ scores whether the Git Agent faithfully tracks the user's evolving intent:
\begin{equation}
R^{\mathbf{track}}_{e,t}
\;=\;
J_\phi\!\left(\widehat{\boldsymbol{Q}}_{i,t},\;\widehat b_{i,t},\;q_{i,t},\;\boldsymbol{I}_{i,t}\right).
\label{eq:tracking_reward}
\end{equation}
The judger evaluates three dimensions: \emph{requirement fidelity} (no omission or hallucination), \emph{base optimality} (maximum compatible reuse), and \emph{staleness exclusion} (withdrawn requirements correctly removed).
These dimensions align with the evaluation metrics in Section~\ref{sec:problem_formulation}, closing the loop between training signal and evaluation protocol.

\paragraph{Training objective.}
For each query we sample a group of $K$~rollouts.
Both reward components are group-normalized into advantages: $A^{\mathbf{out}}_{e}$ across the rollout group and $A^{\mathbf{track}}_{n,t}$ across same-type decisions.
The per-token advantage for Git Agent positions combines them as
\begin{equation}
A_{n,t}
\;=\;
(1-\lambda_{\mathbf{track}})A^{\mathbf{out}}_{e(n)}
\;+\;
\lambda_{\mathbf{track}}\;A^{\mathbf{track}}_{n,t}\,.
\label{eq:combined_advantage}
\end{equation}
We optimize~$\pi_\theta^G$ with a rollout-balanced variant of GRPO~\citep{shao2024deepseekmath}:
\begin{equation}
\mathcal{L}_{\textbf{\textsc{grpo}}}
\;=\;
-\,\frac{1}{K}\sum_{e=1}^{K}
\frac{1}{\lvert\mathcal{I}_e\rvert}
\!\sum_{(n,t)\,\in\,\mathcal{I}_e}\!
\min\!\Big(\rho_{n,t}\,A_{n,t},\;\;\operatorname{clip}\!\left(\rho_{n,t},\;1{-}\epsilon,\;1{+}\epsilon\right)A_{n,t}\Big),
\label{eq:harness_rl_objective}
\end{equation}
where $\mathcal{I}_e$ collects Git Agent output-token positions in rollout~$e$ and $\rho_{n,t}=\exp\!\bigl(\ell_{n,t}(\theta)-\ell_{n,t}^{\mathbf{roll}}\bigr)$ is the importance ratio.
The inner normalization by~$\lvert\mathcal{I}_e\rvert$ ensures that rollouts with richer version-resolution histories do not dominate the gradient update.

%% file: 5_Experiment.tex
\section{Experiments}
\label{sec:experiments}

We evaluate \M~across five domains, two backbones, and three harnesses. We ask: \textbf{RQ1:} Does it improve performance and execution efficiency across models and harnesses? \textbf{RQ2:} Which components drive these gains? \textbf{RQ3:} Can interface-level RL improve the Git Agent, and how do training dynamics and reward balance affect adaptation? \textbf{RQ4:} Can non-parametric guidance distilled from prior inference improve performance without parameter updates? \textbf{RQ5:} Does \M~remain robust as interaction length and task concurrency increase?

\begin{table*}[t]
\centering
\caption{Final-task performance across harnesses. Scores use each benchmark's native metric. Best results are shown in \textbf{bold}, and second-best results are \underline{underlined}.}
\label{tab:main_multiturn}
\fontsize{7.5pt}{8pt}\selectfont
\setlength{\tabcolsep}{2.4pt}
\renewcommand{\arraystretch}{1.22}
\resizebox{\textwidth}{!}{
\begin{tabular}{l l | c c c c c | c c c c c}
\toprule
\rowcolor{gray!30}
\multicolumn{2}{c|}{\textbf{}} &
\multicolumn{5}{c|}{\textbf{Qwen3-32B}} &
\multicolumn{5}{c}{\textbf{DeepSeek-V4-Flash}} \\
\rowcolor{gray!30}
\textbf{Harness} & \textbf{Cross-turn method} &
\textbf{Math} & \textbf{SQL} & \textbf{Search} & \textbf{Code} & \textbf{Research} &
\textbf{Math} & \textbf{SQL} & \textbf{Search} & \textbf{Code} & \textbf{Research} \\
\midrule
\multirow{6}{*}{TCRAG}
& Native             & 72.0 & 53.0 & 1.0 & 3.0 & 25.02 & 74.0 & 69.0 & 34.0 & 56.0 & 42.54 \\
& + Restart           & 73.0 & \underline{64.0} & 2.0 & 5.0 & 16.67 & \underline{88.0} & 66.0 & 41.0 & \underline{61.0} & 43.45 \\
& + U-Fold            & 67.0 & 38.0 & \underline{7.0} & 9.0 & 14.46 & 68.0 & 71.0 & 14.0 & 59.0 & \underline{45.72} \\
& + GCC               & 71.0 & 63.0 & 6.0 & 11.0 & 15.66 & 55.0 & 53.0 & 42.0 & 53.0 & 32.37 \\
\cmidrule{2-12}
\rowcolor[HTML]{F0F6FF}
& + \textbf{\M}      &  \textbf{89.0}  & \textbf{70.0} & \textbf{8.0} & \textbf{19.0} & \textbf{38.45} & \textbf{93.0} & \textbf{80.0} & \textbf{56.0} & \textbf{62.0} & \textbf{46.65}\\
\rowcolor[HTML]{F0F6FF}
& + \textbf{\M-RL} & \underline{84.0} & \underline{64.0} & \underline{7.0} &  \underline{13.0} & \underline{38.02} & 79.0 & \underline{77.0} & \underline{47.0} & 60.0 & 44.78 \\
\midrule
\multirow{6}{*}{StackPlanner}
& Native             & 72.0 & 46.0 & 4.0 & 8.0 & 33.77 & 75.0 & 71.0 & 37.0 & 67.0 & 47.63 \\
& + Restart           & \underline{80.0} & \underline{66.0} & 3.0 & 16.0 & 19.50 & 76.0 & 72.0 & 45.0 & 68.0 & 46.83 \\
& + U-Fold            & 75.0 & 39.0 & 6.0 & 7.0  & 26.24 & 80.0 & \textbf{81.0} & 32.0 & \underline{73.0} & 38.18 \\
& + GCC               & 62.0 & 61.0 & 11.0 & 10.0 & 21.04 & 61.0 & 56.0 & 46.0 & 66.0 & 41.45 \\
\cmidrule{2-12}
\rowcolor[HTML]{F0F6FF}
& + \textbf{\M}      &\textbf{86.0} & \textbf{72.0} & \textbf{15.0} & \underline{17.0} & \underline{38.78} & \textbf{94.0} & \underline{78.0} & \textbf{58.0} & \textbf{74.0} & \textbf{49.96} \\
\rowcolor[HTML]{F0F6FF} 
& + \textbf{\M-RL}  & 77.0 & 65.0 & \underline{12.0} &\textbf{ 18.0} &\textbf{39.20} & \underline{81.0} & 73.0 & \underline{49.0}& 69.0 &\underline{47.83} \\
\midrule
\multirow{6}{*}{OpenHands}
& Native             & 71.0 & 58.0 & 2.0 & 7.0 & 20.22 & 76.0 & 67.0 & 27.0 & 64.0 & 48.88 \\
& + Restart           & 72.0 & 43.0 & 3.0 & 10.0 & 18.20 & 72.0 & 64.0 & 39.0 & 65.0 & \underline{49.76} \\
& + U-Fold            & 47.0 & 22.0 & 2.0 & 6.0 & 8.99 & 71.0 & 75.0 & 16.0 & 59.0 &34.57 \\
& + GCC               & 81.0 & \textbf{77.0} & \underline{10.0} & 11.0 & 13.32 & 62.0 & 63.0 & 38.0 & 65.0 & 39.01 \\
\cmidrule{2-12}
\rowcolor[HTML]{F0F6FF}
& + \textbf{\M}      & \underline{85.0} & \underline{73.0} & \underline{10.0} & \textbf{16.0} & \textbf{39.25} & \textbf{92.0} & \textbf{79.0} & \textbf{57.0} & \textbf{72.0} & \textbf{51.48} \\
\rowcolor[HTML]{F0F6FF}
& + \textbf{\M-RL} & \textbf{86.0} & 63.0 & \textbf{11.0} &  \underline{15.0}& \underline{37.92} & \underline{87.0} & \underline{78.0} & \underline{48.0} & \underline{69.0} & 46.17 \\
\bottomrule
\end{tabular}
}
\end{table*}

\begin{figure*}[t]
    \centering
    \includegraphics[width=\textwidth]{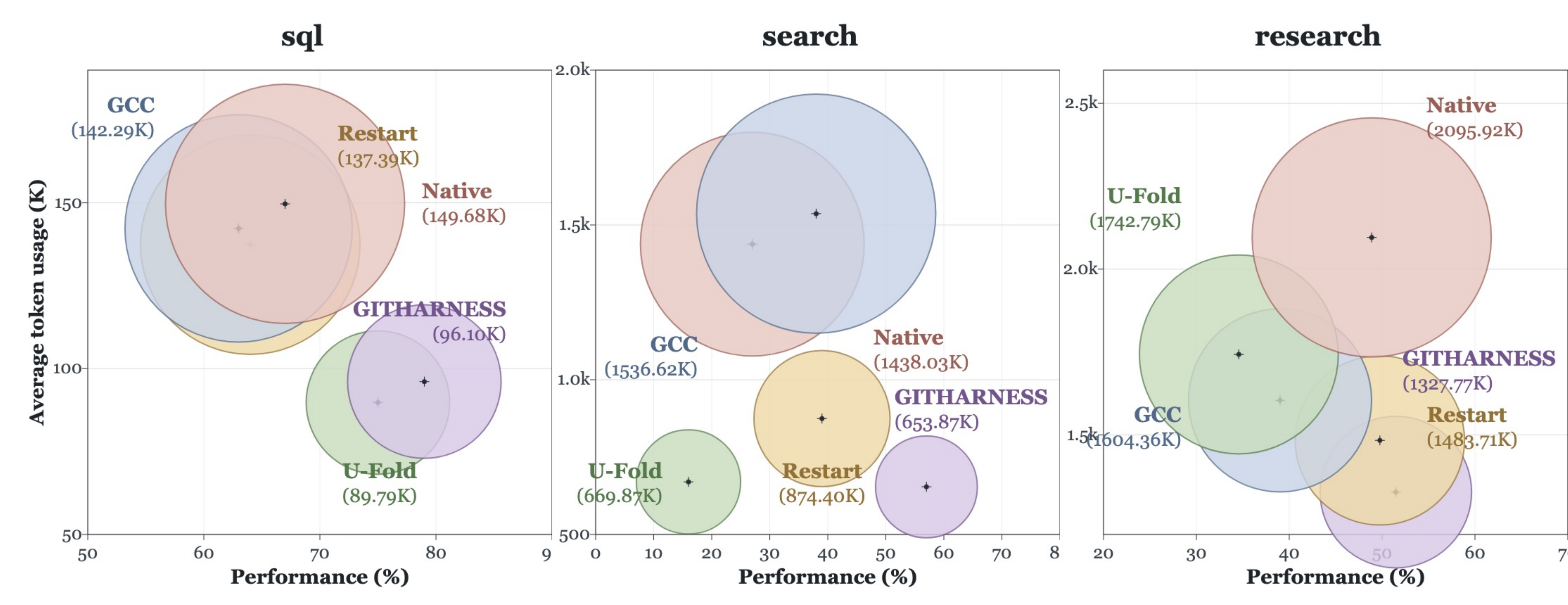}
    \caption{Performance versus token usage under StackPlanner with DeepSeek-V4-Flash. Bubble area denotes estimated API cost; lower right is better.}
    \label{fig:token_performance}
\end{figure*}




\subsection{Experimental Setup}
\label{sec:experimental_setup}

\noindent\textbf{Benchmarks.}
We extend five benchmarks into seven-turn interactions: GSM8K (Math; accuracy), BIRD (SQL; execution accuracy), BrowseComp-Plus (Search; answer accuracy), SWE-bench Verified (Code; resolved rate), and DeepResearch Bench (Research; RACE). Appendix~\ref{appendix:datasets} details the evaluation subsets, trajectory construction, and domain-specific protocols.

\noindent\textbf{Models and harnesses.}
For training-free \M, Qwen3-32B~\citep{yang2025qwen3} or DeepSeek-V4-Flash~\citep{deepseekai2026deepseekv4highlyefficientmilliontoken} serves as both the Git Agent and downstream task executor. For \M-RL, the Git Agent is replaced by a LoRA-GRPO-trained Qwen3-8B, while downstream execution continues to use the model specified by the corresponding experimental setting. We evaluate these configurations with TCRAG (stack memory)~\citep{jiang2024tc}, StackPlanner (hierarchical multi-agent planning)~\citep{zhang2026stackplanner}, and OpenHands (executable workspaces)~\citep{wang2025openhands}. Appendix~\ref{appendix:harnesses} provides additional details on the three harnesses.

\noindent\textbf{Baselines.}
We compare Native, Restart, U-Fold~\citep{su2026u}, and GCC~\citep{wu2025git}, representing native continuation, clean-workspace regeneration, intent-aware context folding, and Git-style context management, respectively. Appendix~\ref{appendix:baselines} details their implementations.

\subsection{Main Results}
\label{sec:main_results}

Table~\ref{tab:main_multiturn} compares methods within each fixed backbone--harness setting; absolute scores are not compared across harnesses because their tools and planning procedures differ.

\noindent\textbf{Overall Effectiveness (RQ1).} Across two backbones, three agent harnesses, and five task domains (Table~\ref{tab:main_multiturn}), \M~outperforms \textit{Native} in all 30 settings and matches or exceeds the strongest competing baseline in 28, including 27 strict wins and one tie. The gains are especially pronounced on Search: with DeepSeek-V4-Flash, \M~improves over the strongest baseline by \textbf{14.0} points under TCRAG and \textbf{18.0} points under OpenHands.

\noindent\textbf{Consistency across Models and Harnesses (RQ1).} Improvements over \textit{Native} are observed with both backbones and all three harnesses. For example, with Qwen3-32B, \M~improves SQL by \textbf{26.0} points under StackPlanner and \textbf{15.0} points under OpenHands; with DeepSeek-V4-Flash, it improves Search by \textbf{22.0} points under TCRAG and \textbf{30.0} points under OpenHands.


\noindent\textbf{Performance and Execution Efficiency (RQ1).} Compared with \textit{Native}, \M~reduces average token usage by \textbf{35.8\%} on SQL, \textbf{54.5\%} on Search, \textbf{73.6\%} on Code, and \textbf{36.7\%} on Research while improving performance in all four domains (Figure~\ref{fig:token_performance} and Table~\ref{tab:token_cost_full}). It also achieves both the best performance and the lowest token usage among all compared methods on Search, Code, and Research. This lower cost is consistent with reusing compatible work and limiting re-execution after local requirement changes.


\subsection{Ablation and Analysis}
\label{sec:ablation_analysis}

\noindent\textbf{Component Ablation (RQ2).} Table~\ref{tab:inference_ablation} shows that both requirement resolution and branch execution contribute to performance and efficiency. Removing \textit{historical version selection} and falling back to
\textit{latest-state resolution}, which fixes the current head as the base
instead of selecting an earlier version, reduces the Code resolved rate from
74.0 to 44.0 while increasing token usage from 1.29M to 4.51M; without
\textsc{Inspect}, the resolved rate falls to 42.0 with 4.22M tokens. Disabling \textsc{ReuseAndPatch} lowers performance and increases token usage across all five domains, while disabling \textsc{FreshSolve} also degrades every task, including drops of \textbf{10.0} points on SQL and \textbf{11.0} points on Search. These results show the value of historical version selection and choosing the execution mode.

\begin{table*}[t]
\centering
\caption{Inference-time ablations under StackPlanner with DeepSeek-V4-Flash. Each task reports its native performance metric and average total token usage (thousands).}
\label{tab:inference_ablation}
\fontsize{7.5pt}{8pt}\selectfont
\setlength{\tabcolsep}{2.4pt}
\renewcommand{\arraystretch}{1.22}
\resizebox{\textwidth}{!}{%
\begin{tabular}{l l | cc cc cc cc cc}
\toprule
\rowcolor{gray!30}
\multicolumn{2}{c|}{}
& \multicolumn{2}{c}{\textbf{Math}}
& \multicolumn{2}{c}{\textbf{SQL}}
& \multicolumn{2}{c}{\textbf{Search}}
& \multicolumn{2}{c}{\textbf{Code}}
& \multicolumn{2}{c}{\textbf{Research}} \\
\rowcolor{gray!30}
\textbf{Component} & \textbf{Variant}
& \textbf{Acc.}$\uparrow$ & \textbf{Tok.}$\downarrow$
& \textbf{EX}$\uparrow$ & \textbf{Tok.}$\downarrow$
& \textbf{EM}$\uparrow$ & \textbf{Tok.}$\downarrow$
& \textbf{Res.}$\uparrow$ & \textbf{Tok.}$\downarrow$
& \textbf{RACE}$\uparrow$ & \textbf{Tok.}$\downarrow$ \\
\midrule
\rowcolor[HTML]{F0F6FF}
\multicolumn{2}{l|}{\textbf{Full GitHarness}}
& 94.0 & 60.0 & 78.0 & 96.1 & 58.0 & 653.9 & 74.0 & 1291.4 & 49.96 & 1327.8 \\
\midrule
\multirow{3}{*}{\shortstack{Requirement\\Resolution}}
& w/o Historical Version Selection   & 84.0 & 104.6 & 62.0 & 142.1 & 42.0 & 1892.6 & 44.0 & 4513.5 & 45.59 & 2299.6 \\
& w/o Inspect                   & 82.0 & 97.6 & 70.0 & 123.4 & 44.0 & 1583.5 & 42.0 & 4215.4 & 43.50 & 2412.3 \\
& w/o Exact-State Reuse         & 88.0 & 107.8 & 76.0 & 130.3 & 54.0 & 963.3 & 63 & 2534.8  & 49.88 & 1360.6 \\
\midrule
\multirow{2}{*}{\shortstack{Branch\\Execution}}
& w/o ReuseAndPatch & 92.0 &65.4  & 75.0 & 104.3 & 53.0 & 862.4 & 69.0 & 1531.4 & 46.63 & 1597.0 \\
& w/o FreshSolve    & 89.0 & 114.2 & 68.0 &102.4 & 47.0 & 793.9 & 72.0 & 1032.5 & 42.56 & 1433.5 \\
\bottomrule
\end{tabular}}
\end{table*}

\begin{figure*}[t]
\centering
\includegraphics[width=\textwidth]{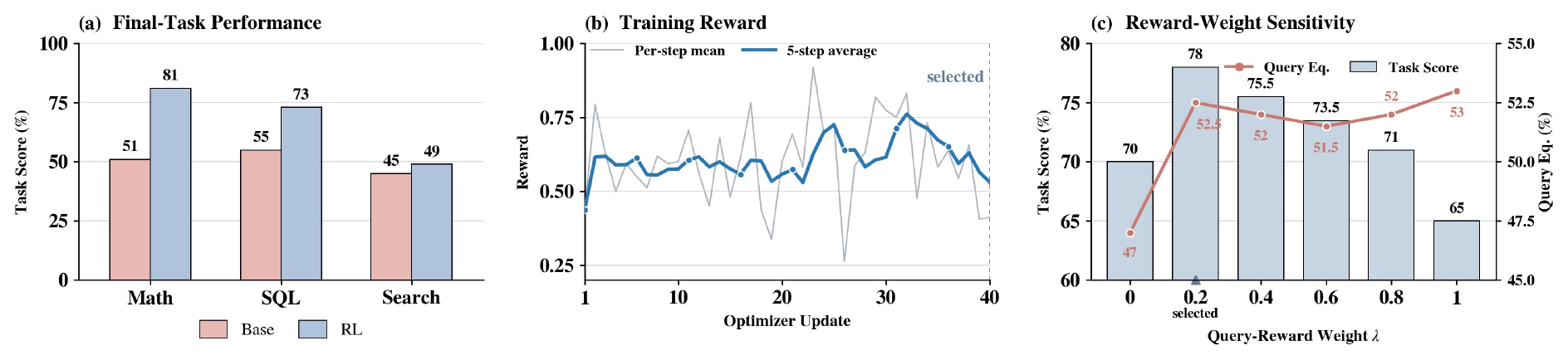}
\caption{Harness RL adaptation with fixed downstream execution: (a) final-task performance on the held-out test set, (b) training reward, and (c) reward-weight sensitivity on the validation set.}
\label{fig:rl_adaptation}
\end{figure*}

\begin{table*}[t]
\centering
\caption{Effect of skill injection under DeepSeek-V4-Flash. Parentheses report changes from the corresponding skill-free GitHarness.}
\label{tab:skill_injection}
\fontsize{8.2pt}{9.2pt}\selectfont
\setlength{\tabcolsep}{2.8pt}
\renewcommand{\arraystretch}{1.18}

\begin{tabular}{@{}l|ccccc|ccccc@{}}
\toprule
\rowcolor{gray!30}
& \multicolumn{5}{c|}{\textbf{StackPlanner}}
& \multicolumn{5}{c}{\textbf{OpenHands}} \\
\rowcolor{gray!30}
\textbf{Method}
& \textbf{Math} & \textbf{SQL} & \textbf{Search} & \textbf{Code} & \textbf{Research}
& \textbf{Math} & \textbf{SQL} & \textbf{Search} & \textbf{Code} & \textbf{Research} \\
\midrule

\M
& 94.0 & 78.0 & \textbf{58.0} & 74.0 & 49.96
& 92.0 & 79.0 & 57.0 & 72.0 & 51.48 \\

\rowcolor[HTML]{F0F6FF}
\textbf{\M-Skills}
& \shortstack{\textbf{95.0}\\{\scriptsize (+1.0)}}
& \shortstack{\textbf{83.0}\\{\scriptsize (+5.0)}}
& \shortstack{\textbf{58.0}\\{\scriptsize (0.0)}}
& \shortstack{\textbf{75.0}\\{\scriptsize (+1.0)}}
& \shortstack{\textbf{52.03}\\{\scriptsize (+2.07)}}
& \shortstack{\textbf{94.0}\\{\scriptsize (+2.0)}}
& \shortstack{\textbf{81.0}\\{\scriptsize (+2.0)}}
& \shortstack{\textbf{58.0}\\{\scriptsize (+1.0)}}
& \shortstack{\textbf{74.0}\\{\scriptsize (+2.0)}}
& \shortstack{\textbf{53.60}\\{\scriptsize (+2.12)}} \\
\bottomrule
\end{tabular}
\end{table*}

\begin{table*}[t]
\centering
\caption{Long-horizon performance under StackPlanner with DeepSeek-V4-Flash.}
\label{tab:long_horizon_results}
\fontsize{7.5pt}{8pt}\selectfont
\setlength{\tabcolsep}{4pt}
\renewcommand{\arraystretch}{1.15}
\begin{tabular}{l|ccccc|ccccc}
\toprule
\rowcolor{gray!30}
& \multicolumn{5}{c|}{\textbf{Math}}
& \multicolumn{5}{c}{\textbf{Research}} \\
\rowcolor{gray!30}
\textbf{Method}
& $H=10$ & $H=15$ & $H=20$ & $H=25$ & $H=30$
& $H=10$ & $H=15$ & $H=20$ & $H=25$ & $H=30$ \\
\midrule
Native
& 73.0 & 72.0 & 69.0 & 63.0 & 62.0
& 42.92 & 41.57 & 37.86 & 35.39 & 34.53 \\
Restart
& 71.0 & 71.0 & 67.0 & 61.0 & 59.0
& 41.05 & 40.08 & 39.14 & 35.88 & 27.70 \\
U-Fold
& 82.0 & 83.0 & 81.0 & 79.0 & 78.0
& 37.53 & 37.13 & 35.41 & 33.52 & 34.16 \\
\rowcolor[HTML]{F0F6FF}
\textbf{\M}
& \textbf{92.0} & \textbf{90.0} & \textbf{87.0} &\textbf{ 85.0} & \textbf{84.0}
& \textbf{49.53} & \textbf{49.27} & \textbf{48.71} & \textbf{47.52} &\textbf{ 47.32} \\
\bottomrule
\end{tabular}
\end{table*}

\noindent\textbf{Harness RL Adaptation (RQ3).} Table~\ref{tab:main_multiturn} reports the frozen test performance of the RL-adapted Git Agent, with the same held-out test comparison shown in Figure~\ref{fig:rl_adaptation}(a) under StackPlanner. With the downstream executor fixed, Harness RL improves the Qwen3-8B Git Agent from 51.0 to 81.0 on Math and from 55.0 to 73.0 on SQL, yielding gains of \textbf{30.0} and \textbf{18.0} points. Although the remaining gap to the training-free DeepSeek Git Agent may partly reflect the smaller 8B policy capacity, the RL-adapted configuration still outperforms the Qwen3-32B configuration on four of five StackPlanner domains.

\noindent\textbf{Training Dynamics and Reward Balance (RQ3).} Figure~\ref{fig:rl_adaptation}(b) reports the per-update training reward and its five-update moving average over 40 updates. Figure~\ref{fig:rl_adaptation}(c) compares query-reward weights on the 200-task validation set: compared with task-only training at $\lambda=0$ (47.0 Query Eq. and 70.0 Task Score), $\lambda=0.2$ improves both metrics. Increasing the weight to $\lambda=1$ raises Query Eq. slightly to 53.0 but reduces Task Score to 65.0, showing that overemphasizing requirement tracking can hurt downstream completion. This result motivates the balanced reward used in the final policy.

\noindent\textbf{Memory-Introduced Guidance (RQ4).} In Table~\ref{tab:skill_injection} and ~\ref{tab:skill_injection_tcrag}, we build a non-parametric guidance memory by distilling reusable strategies from multi-turn inference trajectories on separate source datasets and retrieve relevant guidance at test time. The memory is fixed and collected following SkillRL~\cite{xia2026skillrlevolvingagentsrecursive} when training, and contains no evaluation instances, reference answers, or task-specific evidence; Search, Code, and Research provide no trajectories for its construction. Memory guidance improves 13 of 15 settings, including a \textbf{5.0}-point gain on StackPlanner SQL and Research gains across all three harnesses, ranging from 2.07 to 4.27 points. These results show that experience accumulated during prior inference improves most of the evaluated settings, including held-out tasks, without parameter updates or test leakage.

\noindent\textbf{Performance under Long-Horizon Interleaving (RQ5).} Table~\ref{tab:long_horizon_results} evaluates 10--30-turn sessions interleaving two to four tasks. \M~ranks first at every horizon on both Math and Research. At $H=30$, it exceeds the strongest baseline by 6.0 points on Math and 12.79 RACE points on Research. From $H=10$ to $H=30$, its Research score decreases by only 2.21 points, compared with at least 3.37 points for the baselines, demonstrating robustness to increasing interaction length and task concurrency. Construction details are provided in Appendix~\ref{appendix:datasets}.

%% file: 6_conclusion.tex


\section{Conclusion, Limitations and Future Work}

\M~presents a Git-style version-tracking framework for long-horizon agent collaboration under dynamic user requirements. Instead of mixing evolving requirements and accumulated work in a flat interaction history, \M~binds each effective requirement to its corresponding harness state and organizes them as a branchable version history. Given users' new feedback, a Git Agent reconstructs the current requirement, locates a compatible historical state, and guides the underlying harness to preserve valid work, discard obsolete states, and update only the affected content. Across five task domains, two backbone models, and three agent harnesses, the training-free \M~ configuration outperforms native cross-turn execution in all 30 settings and matches or exceeds the strongest competing baseline in 28 of 30 settings, including 14 of 15 settings under DeepSeek-V4-Flash. Moreover, after interface-level RL adaptation, the 8B Git Agent retains strong performance despite its smaller capacity, outperforming native cross-turn execution in 29 of 30 settings.

For future work, \textsc{GitHarness} depends on accurate requirement interpretation and historical-state selection, which can be challenging under ambiguous or conflicting feedback. Restoring a compatible state also does not guarantee that the underlying agent preserves all unaffected work. Future work will explore uncertainty-aware clarification and finer-grained dependency tracking, and evaluate the framework in longer, real-world interactions.


%% file: 7_appendix.tex
\newpage

\section{Dataset Construction}
\label{appendix:dataset_construction}

\begin{table*}[t]
\centering
\caption{Requirement operators used to construct multi-turn interactions.
Operator labels are hidden from evaluated agents.}
\label{tab:requirement_operators}
\fontsize{8pt}{9pt}\selectfont
\setlength{\tabcolsep}{5pt}
\renewcommand{\arraystretch}{1.12}
\begin{tabular}{lll}
\toprule
\rowcolor{gray!10}
\textbf{Change type} & \textbf{Operator} & \textbf{Effect} \\
\midrule
\multirow{3}{*}{Completion}
& \textsc{Reveal} & Disclose previously unstated source-task information \\
& \textsc{OutputControl} & Revise format, structure, length, or presentation \\
& \textsc{Refine} & Request additional explanation, evidence, or detail \\
\midrule
\multirow{3}{*}{Elicitation}
& \textsc{Constrain} & Add a newly formed executable constraint \\
& \textsc{Comparison} & Add an object, method, metric, or comparison dimension \\
& \textsc{Pivot} & Introduce a related function while retaining compatible context \\
\midrule
\multirow{3}{*}{Shift}
& \textsc{ParameterUpdate} & Replace an active parameter or constraint value \\
& \textsc{Correct} & Correct an active premise, goal, or requirement \\
& \textsc{Withdraw} & Deactivate an active requirement \\
\bottomrule
\end{tabular}
\end{table*}

\paragraph{Final-anchor construction.}
Following the retrospective construction of \citet{tack2026llms}, we use the source task and its verifier as the final anchor:
\begin{equation}
\boldsymbol I_{i,T_i}=\boldsymbol I_i^{\mathrm{src}},
\qquad
V_i=V_i^{\mathrm{src}},
\label{eq:appendix_final_anchor}
\end{equation}
We then construct plausible preceding states that lead to this anchor. Single-turn and multi-turn evaluation therefore share the same final objective and native verifier, isolating the effect of evolving requirements without new answer annotations.

\paragraph{Requirement transition operators.}
We instantiate the three requirement-change directions in Section~\ref{sec:problem_formulation} with nine operators:
\begin{align}
\mathcal O_{\mathrm{Completion}}
&=\{\textsc{Reveal},\textsc{OutputControl},\textsc{Refine}\},\nonumber\\
\mathcal O_{\mathrm{Elicitation}}
&=\{\textsc{Constrain},\textsc{Comparison},\textsc{Pivot}\},\nonumber\\
\mathcal O_{\mathrm{Shift}}
&=\{\textsc{ParameterUpdate},\textsc{Correct},\textsc{Withdraw}\}.
\label{eq:appendix_requirement_operators}
\end{align}

These hidden construction variables are distinct from the Git Agent's \textsc{Inspect}, \textsc{Commit}, and \textsc{Reuse} actions and the update agent's domain tools.

\paragraph{Multi-turn interaction construction.}
Given the final anchor, we construct preceding requirement states and present them chronologically. At turn $t$, the constructor selects a historical anchor $b_{i,t}$, applies an operator $o_{i,t}$, and realizes the change as natural-language feedback $q_{i,t}$:
\begin{equation}
\Delta\boldsymbol I_{i,t}
=o_{i,t}(\boldsymbol I_{i,b_{i,t}}),
\qquad
\boldsymbol I_{i,t}
=\boldsymbol I_{i,b_{i,t}}\oplus\Delta\boldsymbol I_{i,t}.
\label{eq:appendix_requirement_construction}
\end{equation}
Intermediate states may depart from the source task through a transient edit $\delta$ and later recover through a semantic compensating edit $\bar\delta$:
\begin{equation}
(\boldsymbol I\oplus\delta)\oplus\bar\delta
\equiv\boldsymbol I.
\label{eq:appendix_compensating_transition}
\end{equation}
For example, \textsc{Withdraw} can remove a temporary \textsc{OutputControl}, while \textsc{ParameterUpdate} and \textsc{Correct} can restore parameters or premises. A trajectory is retained only if its terminal requirement is semantically equivalent to the source anchor. Construction labels, anchors, and reference states are used only for validation and reward computation and are hidden from evaluated agents.

\section{Additional Method Details}
\label{appendix:method_details}

\paragraph{Inference procedure.}
Algorithm~\ref{alg:githarness_inference} summarizes requirement resolution, branch execution, and version commitment in \M. Normal harness termination and runtime checks do not use the benchmark answer. The task verifier remains restricted to reward computation and offline evaluation.

\begin{algorithm}[t]
\caption{GitHarness inference.}
\label{alg:githarness_inference}
\begin{algorithmic}[1]
\Require Feedback $q_t$, version graph $\mathcal G_{<t}$, checkpoint binding $\mathcal B$, and domain harness $\Gamma_d$
\Ensure Updated work state $\boldsymbol W_t$ or \textsc{Failure}
\State $\eta\leftarrow\varnothing$ and $\xi_t\leftarrow\bot$
\Repeat
    \State $a_k\sim\pi_\theta^G(\cdot\mid q_t,\mathcal G_{<t},\eta)$
    \State $(o_k,\xi_t)\leftarrow\operatorname{Env}_G(\mathcal G_{<t},a_k)$
    \State $\eta\leftarrow\eta\oplus(a_k,o_k)$
\Until{$\xi_t\neq\bot$}
\State Unpack $\xi_t=(\boldsymbol Q_t,b_t,v_t,h_t)$
\If{$\mathcal B(v_t)\neq\bot$}
    \State \Return $\operatorname{Restore}_d(\mathcal B(v_t))$
\EndIf
\If{$b_t=\varnothing$ or $\mathcal B(b_t)=\bot$}
    \State $z_t\leftarrow\textsc{FreshSolve}$
\Else
    \State $z_t\leftarrow R_\psi(\boldsymbol Q_t,\boldsymbol W_{b_t})$
\EndIf
\If{$z_t=\textsc{FreshSolve}$}
    \State $\widetilde{\boldsymbol W}_t^0\leftarrow\operatorname{Empty}_d()$
\Else
    \State $\widetilde{\boldsymbol W}_t^0\leftarrow\operatorname{Fork}_d(\mathcal B(b_t))$
\EndIf
\State $(s_t,\widetilde{\boldsymbol W}_t)\leftarrow\operatorname{Execute}_d(\widetilde{\boldsymbol W}_t^0,\boldsymbol Q_t,h_t)$
\If{$s_t\neq\textsc{Finished}$}
    \State \Return \textsc{Failure}
\EndIf
\State $c_t\leftarrow\operatorname{PersistCheckpoint}_d(\widetilde{\boldsymbol W}_t)$
\State $\mathcal B(v_t)\leftarrow c_t$
\State \Return $\widetilde{\boldsymbol W}_t$
\end{algorithmic}
\end{algorithm}

\paragraph{Update agent.}
For execution modes that require new work, the update agent operates only on an isolated candidate state. Its domain-specific action space is partitioned as
\begin{equation}
\mathcal A_d^U
=
\mathcal A_d^{\mathrm{inspect}}
\cup
\mathcal A_d^{\mathrm{task}}
\cup
\mathcal A_d^{\mathrm{edit}}
\cup
\{\textsc{Finish}\},
\label{eq:appendix_update_agent_actions}
\end{equation}
where inspect actions read recovered state, task actions invoke native domain tools, edit actions update candidate artifacts, and \textsc{Finish} requests normal termination. At update step $j$,
\begin{equation}
u_{t,j}\sim\pi^U
\!\left(\cdot\mid
\boldsymbol Q_t,h_t,z_t,
\widetilde{\boldsymbol W}_{t,j},\zeta_{t,<j}\right),
\qquad
(\widetilde{\boldsymbol W}_{t,j+1},o^U_{t,j})
=\operatorname{Env}_d(\widetilde{\boldsymbol W}_{t,j},u_{t,j}),
\label{eq:appendix_update_agent}
\end{equation}
where $\pi^U$ is the fixed update-agent policy and $\zeta_{t,<j}$ is its within-execution action--observation history. Concrete tools, artifact representations, and stopping protocols remain domain specific. \textsc{ReuseAndPatch} denotes continued execution from a fork rather than a deterministic file patch, while \textsc{FreshSolve} and \textsc{ReuseAndPatch} are runtime initialization decisions rather than update-agent editing actions.

\begin{table}[t]
\centering
\caption{Semantic action classes available to the update agent.}
\label{tab:update_agent_actions}
\fontsize{8pt}{9pt}\selectfont
\setlength{\tabcolsep}{4pt}
\renewcommand{\arraystretch}{1.12}
\begin{tabular}{lll}
\toprule
\rowcolor{gray!10}
\textbf{Class} & \textbf{Generic operations} & \textbf{Purpose} \\
\midrule
Inspect & \textsc{Glob}, \textsc{Grep}, \textsc{Read} & Inspect files or recovered work units \\
Task & Domain tools & Perform domain-specific task operations \\
Edit & \textsc{Write}, \textsc{Patch}, \textsc{Delete} & Modify only the isolated candidate state \\
Finish & \textsc{Finish} & Return the result and request runtime finalization \\
\bottomrule
\end{tabular}
\end{table}

\paragraph{Harness RL rollout extraction.}
An episode contains interleaved calls to the Git Agent, version runtime, update agent, and domain tools. For interface call $n$, we record $\xi_{e,n}=(X_{e,n},Y_{e,n},\boldsymbol\ell^{\mathrm{roll}}_{e,n})$, where $X_{e,n}$ is the assembled context, $Y_{e,n}$ the sampled response, and $\boldsymbol\ell^{\mathrm{roll}}_{e,n}$ its rollout log-probabilities. The training mask is
\begin{equation}
m_{e,n,j}
=
\mathbb I\!\left[
\xi_{e,n}\ \text{is a Git-Agent call and token }j\in Y_{e,n}
\right],
\qquad
\mathcal I_e=\{(n,j):m_{e,n,j}=1\}.
\label{eq:appendix_harness_rl_mask}
\end{equation}
Only positions in $\mathcal I_e$ contribute to the policy update. User, system, runtime, tool-observation, and update-agent tokens remain in the rollout context but receive no gradient. Episode rewards are assigned to the corresponding Git-Agent segments before group normalization and the GRPO update defined in Section~\ref{sec:harness_rl_training}.

\section{Experiment Datasets}
\label{appendix:datasets}

We evaluate \M~on five benchmarks spanning mathematical reasoning, text-to-SQL generation, web search, software engineering, and deep research. Only Math and SQL provide training trajectories for RL adaptation; Search, Code, and Research are held out entirely from RL training and serve as out-of-domain transfer evaluations. The Math and SQL candidate pools contain 400 tasks in total: 200 are used for RL training and 200 are reserved for validation. Table~\ref{tab:dataset_overview} summarizes the task-disjoint training, validation, and test splits.

\paragraph{\ding{182} GSM8K~\citep{cobbe2021trainingverifierssolvemath} (Math).}
GSM8K contains 8.5K linguistically diverse grade-school math problems requiring multi-step arithmetic reasoning. From 200 candidate tasks, we use 100 for RL training and hold out 100 for validation; a separate 100-task test set is used for final evaluation. We evaluate the final answer using normalized exact-match accuracy, where extracted numerical answers are normalized before comparison with the gold answer.

\paragraph{\ding{183} BIRD~\citep{li2023can} (SQL).}
BIRD is a large-scale, database-grounded text-to-SQL benchmark containing 12,751 question--query pairs across 95 databases and 37 domains. From 200 candidate tasks, we use 100 for RL training and hold out 100 for validation; a separate 100-task test set is used for final evaluation. We report execution accuracy by executing the predicted and gold SQL queries on the same SQLite database and comparing their returned results; gold SQL and execution results remain evaluator-only.

\paragraph{\ding{184} BrowseComp-Plus~\citep{chen2025browsecomp} (Search).}
BrowseComp-Plus extends BrowseComp with a fixed, human-verified document corpus designed to provide controlled conditions for evaluating search agents. We use 100 queries and evaluate the final answers using normalized exact-match accuracy. Citation validity is additionally recorded as diagnostic information but is not used to determine the primary accuracy score.

\paragraph{\ding{185} SWE-bench Verified~\citep{jimenez2024swe} (Code).}
SWE-bench Verified contains 500 human-validated software engineering tasks derived from real GitHub issues. We construct a fixed subset of 100 tasks and evaluate the resulting patches in isolated environments using the official SWE-bench evaluation harness, with the resolved outcome as the success criterion.

\paragraph{\ding{186} DeepResearch Bench~\citep{du2025deepresearch} (Research).}
DeepResearch Bench contains 100 PhD-level research tasks spanning 22 expert-curated domains, with 50 English and 50 Chinese queries. We use all 100 tasks and evaluate generated reports using RACE, which measures report quality against the benchmark's reference reports under its predefined evaluation criteria.

\begin{table}[t]
\centering
\small
\caption{Task-disjoint training, validation, and test subsets. Only the training trajectories are used for RL adaptation.}
\label{tab:dataset_overview}
\setlength{\tabcolsep}{5pt}
\renewcommand{\arraystretch}{1.08}
\begin{tabular}{@{}lcccc@{}}
\toprule
\rowcolor{gray!10}
\textbf{Dataset} & \textbf{RL Train} & \textbf{Validation} & \textbf{Test} & \textbf{Lang.} \\
\midrule
GSM8K              & 100 & 100 & 100 & EN    \\
BIRD                & 100 & 100 & 100 & EN    \\
BrowseComp-Plus     & 0   & 0   & 100 & EN    \\
SWE-bench Verified  & 0   & 0   & 100 & EN    \\
DeepResearch Bench  & 0   & 0   & 100 & EN/ZH \\
\bottomrule
\end{tabular}
\end{table}

\paragraph{\ding{187} Multi-turn interaction construction.}
For each source task, we construct a multi-turn interaction sequence using three controlled transition classes. \textsc{Preserve} retains the current goal and active requirements while revealing compatible information, changing presentation, or requesting further explanation. \textsc{Extend} keeps the current task active while adding a constraint, comparison dimension, or related task function. \textsc{Revise} corrects, replaces, withdraws, or restores an active requirement. By construction, the final effective query is semantically equivalent to the original source task, allowing the final artifact to be evaluated with the benchmark's original evaluator.

\paragraph{\ding{188} Interleaved long-horizon conversations.}
To evaluate multi-requirement tracking and cross-task interference, we compose multiple task-disjoint seven-turn interaction sequences into a single conversation with $H\in\{10,15,20,25,30\}$ user turns. The conversation interleaves two to four active tasks, and each turn may update one task or jointly update several tasks. When needed to reach the target length, we insert continuation or progress-check requests (e.g., ``continue working'' or ``report the current progress'') that leave the active requirements unchanged. Hidden task states and transition labels remain unavailable to the agent. The final turn requests the outputs of all active tasks, which are verified separately using their benchmark-specific evaluators and then macro-averaged.

\section{Harness and Adapter Details}
\label{appendix:harnesses}

We evaluate all methods with three agent harnesses: TC-RAG, StackPlanner, and OpenHands. Each harness retains its native reasoning and tool-use procedure, while a lightweight adapter connects it to the corresponding benchmark interface.

\paragraph{\ding{182} TC-RAG.}
TC-RAG \cite{jiang2024tc} employs a memory-stack mechanism for adaptive retrieval, reasoning, and planning. Through Push and Pop operations, it manages intermediate states, controls the retrieval process, and mitigates the accumulation of erroneous knowledge.

\paragraph{\ding{183} StackPlanner.}
StackPlanner \cite{zhang2026stackplanner} is a centralized hierarchical multi-agent framework that decouples high-level coordination from subtask execution. It employs active task-level memory control and structured experience memory to support long-horizon coordination and the reuse of prior coordination experience.

\paragraph{\ding{184} OpenHands.}
OpenHands \citep{wang2025openhands} is a general-purpose platform for developing AI agents that interact with their environment by writing code, operating through a command line, and browsing the web. It also provides sandboxed environments for safe code execution.

\section{Baseline Implementation Details}
\label{appendix:baselines}

We compare \M~with four baselines that adopt different strategies for cross-turn context management: native state retention, context reconstruction, context compression, and versioned context management.

\paragraph{\ding{182} Native.}
Native directly relies on the conversational state and workspace maintained by the underlying task harness, without introducing any additional cross-turn context-management module. Information from previous turns is therefore retained only through the harness's native state mechanism.

\paragraph{\ding{183} Restart.}
Restart reconstructs the accumulated user requirements at each turn into a self-contained query and executes it in a fresh harness session with a clean workspace. Previous reasoning traces and intermediate artifacts are discarded, so each turn is solved independently based on the reconstructed requirement.

\paragraph{\ding{184} U-Fold.}
U-Fold \citep{su2026u} retains the full user--agent dialogue and tool-call history while constructing a compact working context at each turn. It maintains an intent-aware, evolving dialogue summary together with a compact, task-relevant tool log, which are used to support the current interaction.

\paragraph{\ding{185} GCC.}
GCC \citep{wu2025git} organizes agent context as a version-controlled persistent file system. It provides four explicit operations---COMMIT, BRANCH, MERGE, and CONTEXT---to checkpoint progress, explore alternative reasoning paths, integrate branch states, and retrieve historical context at different levels of granularity.

\section{Implementation Details}
\label{appendix:implementation}

\paragraph{Domain-specific execution environments.}
We instantiate the downstream update agent with the capabilities required by each benchmark while leaving requirement resolution unchanged. Table~\ref{tab:implementation_domain_tools} summarizes the model-visible tools, persistent work products, and final verification for each domain. Within each backbone--harness block, all compared methods use the same downstream model, task inputs, model-visible tools, execution budget, and verifier. Across all domains, gold annotations, hidden evaluation resources, and verifier feedback remain evaluator-only; method-specific persistent state remains part of the system being evaluated.

\begin{table*}[t]
\centering
\caption{Domain-specific tools and persistent work products.}
\label{tab:implementation_domain_tools}
\small
\setlength{\tabcolsep}{3.8pt}
\renewcommand{\arraystretch}{1.12}
\begin{tabular}{@{}
>{\raggedright\arraybackslash}p{0.07\textwidth}
>{\raggedright\arraybackslash}p{0.50\textwidth}
>{\raggedright\arraybackslash}p{0.38\textwidth}@{}}
\toprule
\rowcolor[HTML]{F2F2F2}
\textbf{Domain} & \textbf{Model-visible tools and constraints} & \textbf{Persistent work products} \\
\midrule
Math
& Deterministic calculator and scoped workspace read/write/patch operations
& \texttt{answer.md} and calculation evidence \\

SQL
& SQL generation from the public schema and BIRD evidence; no database access during inference
& \texttt{answer.sql} \\

Search
& Read-only BrowseComp retrieval returning top-5 passages with 512-token snippets
& \texttt{answer.json} and runtime-recorded retrieval evidence \\

Research
& Bounded read-only BoCha web search
& \texttt{report.md}, report outline, and runtime-recorded retrieval evidence \\

Code
& Repository inspection, editing, shell execution, and repository-local test execution in an isolated workspace; no access to hidden tests
& Repository patch \\
\bottomrule
\end{tabular}
\end{table*}

\paragraph{Reinforcement-learning adaptation.}
We train only the Qwen3-8B Git Agent with LoRA GRPO. During RL training, the fixed downstream executor and query-equivalence judge use DeepSeek-V4-Flash, and all non-Git-Agent components remain frozen. Each optimizer update contains four prompt groups, with two Math and two SQL cases, and samples eight independent rollouts per group. Advantages are normalized only among rollouts generated from the same prompt. The terminal episode score is assigned to every trainable Git-Agent segment in that episode, while tokens produced by the runtime, downstream executor, tools, and verifiers are excluded from the loss. The selected checkpoint is trained for 40 updates, corresponding to 160 prompt groups and 1,280 sampled episodes. The reward weight $\lambda=0.2$ is selected on the 200-task validation set described in Appendix~\ref{appendix:datasets}. Table~\ref{tab:rl_training_configuration} reports the complete training configuration.

\paragraph{Long-horizon finalization.} When the final message requests outputs for several active tasks, the experiment scheduler extracts one task-local request for each  referenced task and invokes the same GitHarness procedure independently on its version history and workspace. Each invocation produces one terminal \textsc{Commit} or  \textsc{Reuse}, after which the scheduler aggregates the task outputs into one response. Thus, a global final message may trigger several task-local terminal actions without changing the single-handoff semantics of GitHarness.

\begin{table*}[t]
\centering
\caption{LoRA-GRPO training configuration for the Git Agent.}
\label{tab:rl_training_configuration}
\footnotesize
\setlength{\tabcolsep}{4pt}
\renewcommand{\arraystretch}{1.14}
\begin{tabular}{@{}
>{\raggedright\arraybackslash}p{0.18\textwidth}
>{\raggedright\arraybackslash}p{0.27\textwidth}
>{\raggedright\arraybackslash}p{0.18\textwidth}
>{\raggedright\arraybackslash}p{0.27\textwidth}@{}}
\toprule
\rowcolor[HTML]{F2F2F2}
\textbf{Parameter} & \textbf{Value} & \textbf{Parameter} & \textbf{Value} \\
\midrule
\rowcolor[HTML]{F7F7F7}
\multicolumn{4}{@{}l}{\textbf{Models and execution}} \\
Trainable policy & Qwen3-8B Git Agent
& Downstream executor model & DeepSeek-V4-Flash \\
Query-equivalence judge & DeepSeek-V4-Flash
& Adaptation & LoRA \\
Git-Agent tool steps & 4
& Downstream tool steps & 16 \\
Git-Agent context / response limit & 16,384 / 4,096 tokens
& Downstream response limit & 4,096 tokens \\
\midrule
\rowcolor[HTML]{F7F7F7}
\multicolumn{4}{@{}l}{\textbf{Sampling}} \\
Updates & 40
& Checkpoint interval & Every update \\
Prompt groups/update & 4 (2 Math + 2 SQL)
& Rollouts/group & 8 \\
Sampled episodes & 1,280
& Temperature / top-$p$ & 0.8 / 1.0 \\
\midrule
\rowcolor[HTML]{F7F7F7}
\multicolumn{4}{@{}l}{\textbf{Optimization}} \\
LoRA rank / alpha & 16 / 32
& LoRA dropout & 0 \\
Target modules & Q/K/V/O and MLP projections
& Precision & bfloat16 \\
Optimizer & AdamW
& Learning rate & $2\times10^{-6}$ \\
Adam betas / weight decay & $(0.9,0.95)$ / 0
& Schedule & Constant \\
GRPO clip range & 0.1
& Maximum gradient norm & 1.0 \\
Base-policy KL coefficient & 0.01
& Query-reward weight $\lambda$ & 0.2 \\
\bottomrule
\end{tabular}
\end{table*}

\section{Additional Experiments}
\label{sec:additional-experiments}
\paragraph{\ding{182} Efficiency and Cost Details.}
\label{appendix:efficiency}

\begin{figure*}[t]
\centering
\includegraphics[width=0.65\textwidth]{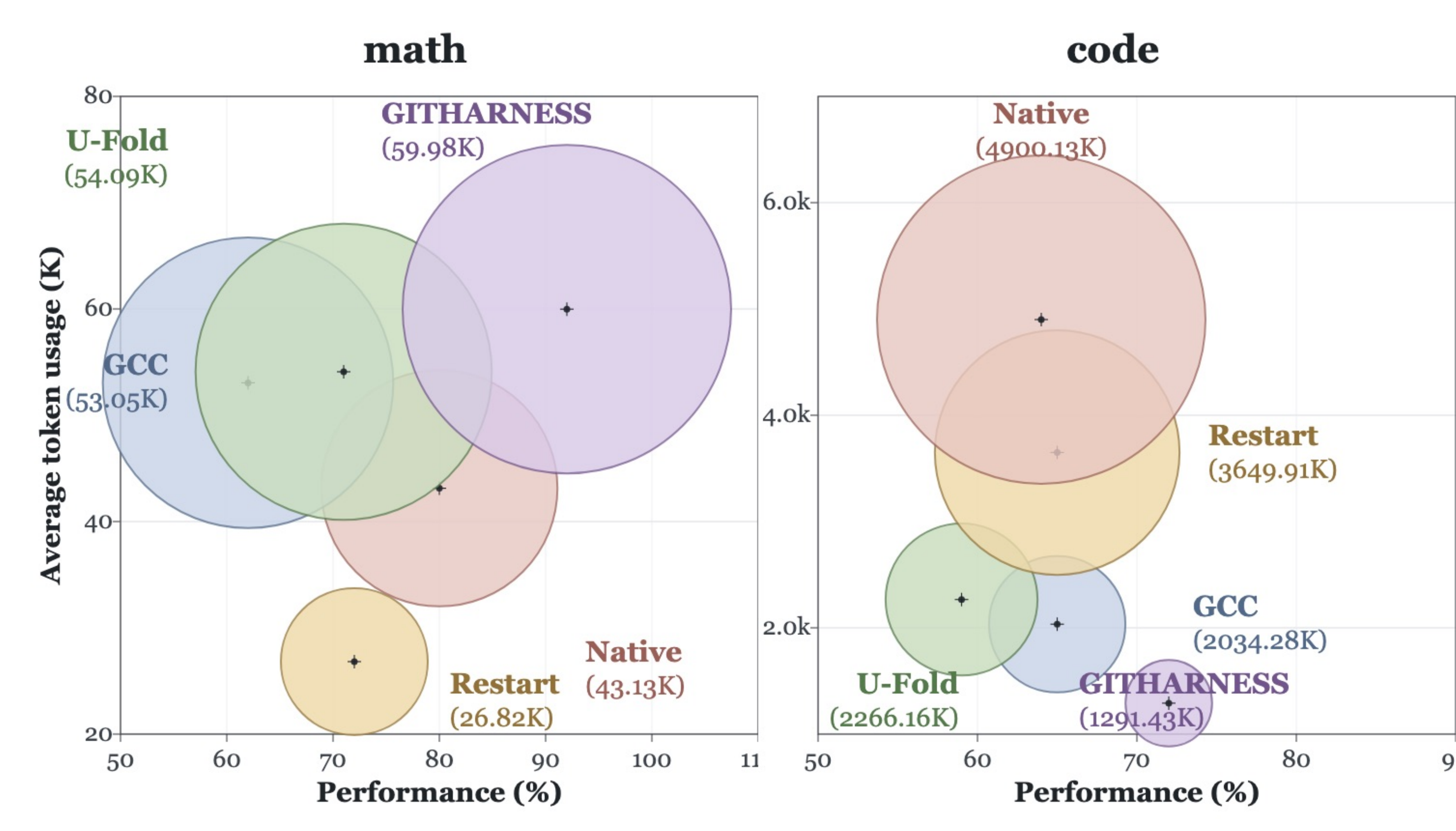}
\caption{Performance versus token usage on Math and Code under StackPlanner with DeepSeek-V4-Flash. Bubble area denotes estimated API cost; lower right is better.}
\label{fig:token_performance_remaining}
\end{figure*}

\begin{figure*}[t]
\centering
\includegraphics[width=\textwidth]{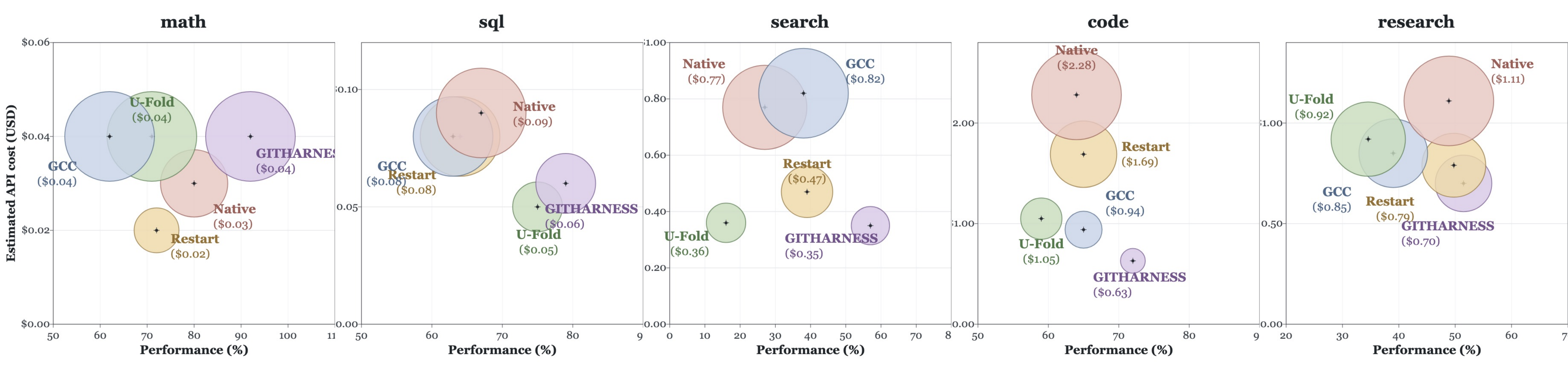}
\caption{Performance versus estimated API cost across five domains under StackPlanner with DeepSeek-V4-Flash. Bubble area denotes average token usage; lower right is better.}
\label{fig:cost_performance}
\end{figure*}

Figures~\ref{fig:token_performance_remaining} and~\ref{fig:cost_performance} complement the main-text efficiency analysis. The former reports the Math and Code token--performance trade-offs omitted from Figure~\ref{fig:token_performance} for readability, while the latter compares estimated API cost across all five domains. We aggregate all model calls over each seven-turn interaction and average over completed samples. \M~achieves the lowest token usage and estimated cost on Search, Code, and Research; Math and SQL exhibit task-dependent trade-offs. Table~\ref{tab:token_cost_full} provides the corresponding numerical results.

\begin{table*}[t]
\centering
\caption{Detailed token usage (thousands) and API cost (USD) under StackPlanner with DeepSeek-V4-Flash.}
\label{tab:token_cost_full}
\fontsize{7.5pt}{8pt}\selectfont
\setlength{\tabcolsep}{2.8pt}
\renewcommand{\arraystretch}{1.12}
\resizebox{\textwidth}{!}{%
\begin{tabular}{l | cc cc cc cc cc}
\toprule
\rowcolor{gray!30}
& \multicolumn{2}{c}{\textbf{Math}}
& \multicolumn{2}{c}{\textbf{SQL}}
& \multicolumn{2}{c}{\textbf{Search}}
& \multicolumn{2}{c}{\textbf{Code}}
& \multicolumn{2}{c}{\textbf{Research}} \\
\rowcolor{gray!30}
\textbf{Method}
& \textbf{Tokens} & \textbf{Cost}
& \textbf{Tokens} & \textbf{Cost}
& \textbf{Tokens} & \textbf{Cost}
& \textbf{Tokens} & \textbf{Cost}
& \textbf{Tokens} & \textbf{Cost} \\
\midrule
Native  & 43.13 & 0.03 & 149.68 & 0.09 & 1,438.03 & 0.77 & 4,900.13 & 2.28 & 2,095.92 & 1.11 \\
Restart & 26.82 & 0.02 & 137.39 & 0.08 &   874.40 & 0.47 & 3,649.91 & 1.69 & 1,483.71 & 0.79 \\
U-Fold  & 54.09 & 0.04 &  89.79 & 0.05 &   669.87 & 0.36 & 2,266.16 & 1.05 & 1,742.79 & 0.92 \\
GCC     & 53.05 & 0.04 & 142.29 & 0.08 & 1,536.62 & 0.82 & 2,034.28 & 0.94 & 1,604.36 & 0.85 \\
\midrule
\rowcolor[HTML]{F0F6FF}
\textbf{\M}
& 59.98 & 0.04 & 96.10 & 0.06 & 653.87 & 0.35 & 1,291.43 & 0.63 & 1,327.77 & 0.70 \\
\bottomrule
\end{tabular}}
\end{table*}

\begin{figure*}[t]
\centering
\includegraphics[width=\textwidth]{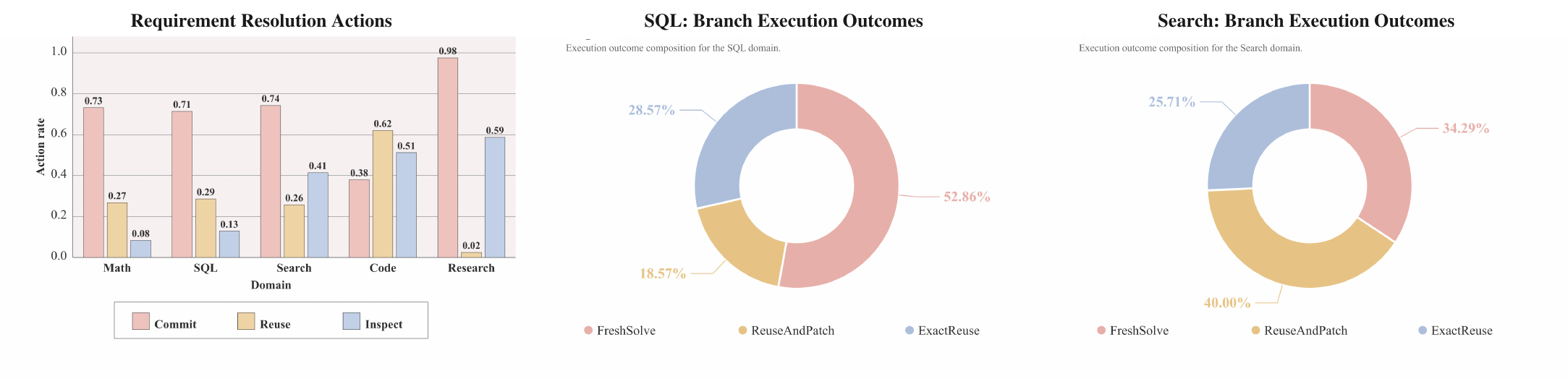}
\caption{Requirement-resolution actions across all domains and branch-execution strategies for SQL and Search.}
\label{fig:resolution_sql_search}
\end{figure*}

\begin{figure*}[t]
\centering
\includegraphics[width=\textwidth]{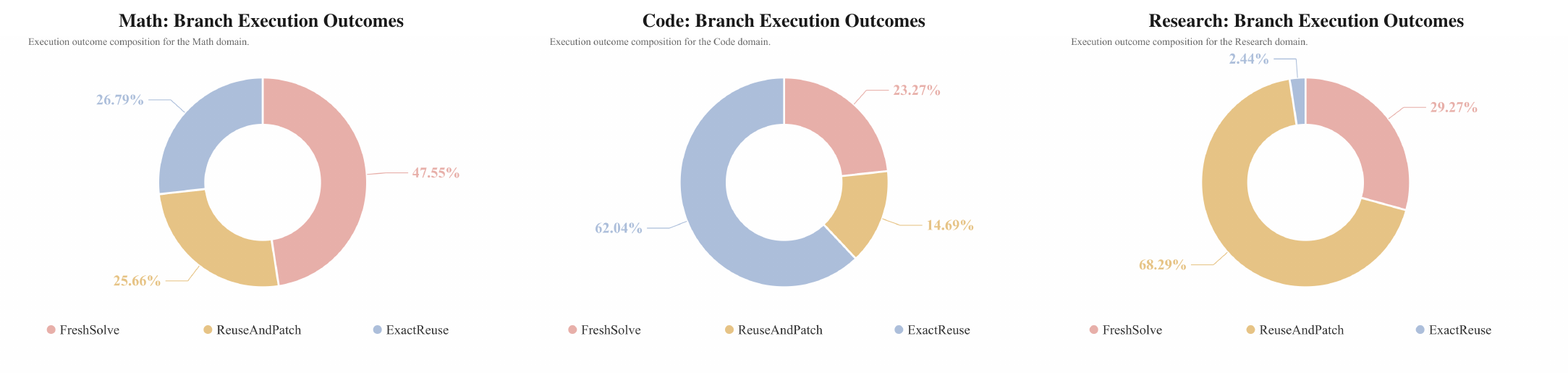}
\caption{Branch-execution strategies for Math, Code, and Research.}
\label{fig:branch_execution_remaining}
\end{figure*}

\begin{table}[t]
\centering
\caption{Effect of skill injection with TCRAG under DeepSeek-V4-Flash.}
\label{tab:skill_injection_tcrag}
\fontsize{8pt}{9pt}\selectfont
\setlength{\tabcolsep}{3.5pt}
\renewcommand{\arraystretch}{1.16}
\begin{tabular}{l|ccccc}
\toprule
\rowcolor{gray!30}
\textbf{Method}
& \textbf{Math} & \textbf{SQL} & \textbf{Search}
& \textbf{Code} & \textbf{Research} \\
\midrule
\M
& 93.0 & 80.0 & \textbf{56.0} & 62.0 & 46.65 \\
\rowcolor[HTML]{F0F6FF}
\textbf{\M-Skills}
& \shortstack{\textbf{96.0}\\{\scriptsize (+3.0)}}
& \shortstack{\textbf{81.0}\\{\scriptsize (+1.0)}}
& \shortstack{55.0\\{\scriptsize (-1.0)}}
& \shortstack{\textbf{63.0}\\{\scriptsize (+1.0)}}
& \shortstack{\textbf{50.92}\\{\scriptsize (+4.27)}} \\
\bottomrule
\end{tabular}
\end{table}

\paragraph{\ding{183} Decision Behavior Analysis.}
\label{appendix:decision_behavior}

Figures~\ref{fig:resolution_sql_search} and~\ref{fig:branch_execution_remaining} show that \M does not collapse to a single resolution action or execution strategy. During requirement resolution, the Git Agent most frequently selects \textsc{Commit} in Math, SQL, Search, and Research, whereas Code more often reuses an existing requirement version; \textsc{Inspect} is particularly common in Code and Research. During branch execution, \textsc{FreshSolve} is most frequent in Math and SQL, \textsc{ReuseAndPatch} dominates Research and is common in Search, and \textsc{ExactReuse} dominates Code. These domain-dependent patterns show that \M~adapts both its historical-version selection and downstream execution strategy to the current update.

\paragraph{\ding{184} Additional Skill-Injection Results.}
\label{appendix:skill_injection}

Table~\ref{tab:skill_injection_tcrag} complements the main analysis with results under TCRAG, a harness based on stack-structured task memory. Skill injection improves four of the five domains, including gains of $3.0$ points on Math and $4.27$ points on Research. Together with the StackPlanner and OpenHands results in Table~\ref{tab:skill_injection}, these results show that reusable version-decision experience transfers across harnesses with different state-management mechanisms.




\section{Computational Resources and Software Environment}
\noindent\textbf{Compute.}
Most inference experiments ran on a shared Ubuntu~22.04.5 LTS server with two Intel Xeon Platinum~8358 CPUs, 1007~GiB of RAM, and eight NVIDIA A800-SXM4-80GB GPUs. Harness-RL training was conducted separately on one NVIDIA H800 GPU. 

\noindent\textbf{Model serving.}
Qwen3-32B was served through an OpenAI-compatible vLLM~0.19.0 endpoint using two A800 GPUs, tensor parallelism of 2, bfloat16 precision, and a 61,440-token context limit. DeepSeek experiments used the official \texttt{deepseek-v4-flash} API with its default thinking configuration (enabled with high reasoning effort) and a 1M-token context window. All cross-turn methods within a backbone--harness--task cell used the same reasoning configuration, generation limits, and agent-step budget.

\noindent\textbf{Task environments.}
BIRD predictions were executed against the benchmark's local SQLite databases and scored by execution accuracy. BrowseComp-Plus used its fixed 100,195-document corpus with Qwen3-Embedding-8B and FAISS retrieval on GPU; each search returned five passages truncated to 512 tokens. SWE-bench Verified instances ran in dedicated Docker~29.1.3 containers and were graded with the official harness. Research reports were evaluated with the official DeepResearch Bench pipeline, using GPT-5.5 for RACE.

\noindent\textbf{Cost accounting.}
We report model-token cost per completed sample using peak cache-miss pricing from September~14, 2026: \$0.44 per million input tokens and \$1.32 per million output tokens.






\IfFileExists{figs/prompt.tex}{%
\newpage


\input{figs/prompt}

}{}

%% file: figs/prompt.tex
\section{Prompt Templates}
\label{appendix:prompts}

\begin{table*}[t]
\centering
\caption{Values used for the domain-specific semantic-check placeholder.}
\label{tab:query_equivalence_domain_checks}
\small
\setlength{\tabcolsep}{4pt}
\renewcommand{\arraystretch}{1.10}
\begin{tabular}{@{}
>{\raggedright\arraybackslash}p{0.10\textwidth}
>{\raggedright\arraybackslash}p{0.84\textwidth}@{}}
\toprule
\rowcolor[HTML]{F2F2F2}
\textbf{Domain} & \textbf{Additional semantic check} \\
\midrule
Math & Preserve the mathematical goal, givens, variables, units,
relationships, active constraints, and requested output. \\
SQL & Preserve the database question, requested columns or aggregation,
filters, grouping, ordering, limits, comparisons, and aliases. Treat fixed
schema and SQL transport wrappers as harness context. \\
Search & Preserve the information need, entities, disambiguating clues, time
range, requested answer granularity, and evidence or citation requirements. \\
Research & Preserve the research question, scope, subjects, time and region
bounds, requested synthesis, deliverable structure, evidence, and citation
requirements. Do not reduce the task to short-answer search. \\
Code & Preserve the software issue goal, repository-scoped entities, affected
symbols or files, reproduction and failure behavior, expected behavior, version
constraints, requested change scope, and active testing or output requirements. \\
\bottomrule
\end{tabular}
\end{table*}
This section presents the core prompts used by \M. Because some prompts are lengthy, we omit repeated low-level instructions and examples for presentation; the released code provides the complete prompts used in all experiments.

\subsection{Requirement Resolution}
The Git Agent uses the following prompt to convert each user turn into a complete resolved requirement and to choose among \textsc{Inspect}, \textsc{Commit}, and \textsc{Reuse}. The box is an abridged presentation rather than a verbatim serialization; the implementation terms \texttt{state\_forest} and \texttt{resolved\_query} correspond to the paper's version graph and resolved requirement, respectively.
\begin{promptbox}{Git Agent Requirement-Resolution Prompt (Abridged)}
You track the user's current requirement state over a version graph.

Use only the provided version-graph tools. Do not answer or execute the task.
\textbf{Critical update procedure:}

1. Unless \texttt{current\_request}
explicitly withdraws or restores content, use
\texttt{active\_state.resolved\_query} as the mandatory source. Edits are
cumulative. Do not inspect or base an ordinary correction on an ancestor;
reuse an ancestor only when its complete requirement exactly matches the
desired requirement.

2. Treat entities, attributes, numbers, relationships, filters, projections,
and outcomes as semantic slots. Replace every stale occurrence of a corrected
slot, including evidence and question wording; preserve other slots and their
required dependencies. Never leave conflicting values for one slot.

\textit{Steps 3--4, the context-schema explanation, repeated validation
rules, formatting sanitization, and illustrative examples are omitted here
for space.}

5. Before \textsc{Commit} or \textsc{Reuse}, verify the latest requested
outcome and output specification, the latest slot values and dependencies, all
unmentioned active requirements, and the absence of withdrawn content. If any
check fails, do not issue a terminal action.

Every \texttt{resolved\_query} must be the complete, self-contained current
requirement, never a delta such as ``change X and keep everything else.''
Content fidelity has priority over brevity. Silently distinguish the stable
requirement core, persistent updates, temporary constraints, and presentation
requirements such as precision, tables, or output format.

Apply this decision priority: explicit withdrawal or restoration, explicit
goal change, persistent local edit, temporary addition, and exact no-op. A
higher-priority instruction overrides lower-priority preservation rules. For
an ordinary local edit, apply only the requested change to the active resolved
requirement and preserve every unmentioned requirement-bearing clause. For an
explicit goal change, rebuild the complete requirement around the new requested
outcome while retaining accepted facts needed for that outcome and dropping
content that served only the previous goal.

\textit{Detailed rules for comparison construction, restoration-scope edge
cases, formatting sanitization, and illustrative examples are omitted here for
space.}

For explicit withdrawal or restoration, use the version index and parent links
to identify the historical version immediately preceding the withdrawn content,
inspect that candidate, and replay only later edits that remain active. If the
desired requirement exactly matches an inspected version, call
\texttt{reuse\_resolved\_state}; otherwise call
\texttt{commit\_resolved\_query} with the appropriate base version.
\texttt{commit\_resolved\_query} and \texttt{reuse\_resolved\_state} are
terminal on success.

\textbf{Runtime input:} \texttt{current\_request},
\texttt{active\_state}, and \texttt{history\_index}.

\textbf{Output:} Return exactly one structured tool call. Inspection is
non-terminal; commit and reuse are terminal. Fields marked optional may be
omitted, and no additional fields are allowed.

\noindent\texttt{inspect\_state\_forest}\par
\noindent\texttt{\{"tree\_id":"<string, optional>",}\par
\noindent\texttt{"state\_id":"<string, optional>"\}}

\noindent\texttt{commit\_resolved\_query}\par
\noindent\texttt{\{"base\_state\_id":"<string or null>",}\par
\noindent\texttt{"resolved\_query":"<complete requirement>",}\par
\noindent\texttt{"summary":"<short change summary>"\}}

\noindent\texttt{reuse\_resolved\_state}\par
\noindent\texttt{\{"state\_id":"<string>",}\par
\noindent\texttt{"summary":"<short reuse summary>"\}}
\end{promptbox}

\subsection{Downstream Execution Routing}
When a verified workspace is available, the following prompt decides whether the downstream executor should solve from scratch, reuse and patch the workspace, or inspect it before deciding. The \texttt{change\_hint} is a non-authoritative summary generated by the Git Agent from model-visible interaction history; it contains no evaluator reference, gold answer, or gold edit.

\begin{promptbox}{Downstream Execution Router Prompt}
You choose how the downstream executor should complete a new resolved task when a verified base workspace is available. Use exactly one routing tool per response.

The \texttt{current\_resolved\_task} is authoritative. The \texttt{base\_resolved\_task} describes what the existing workspace was built for. \texttt{change\_hint} is only a hint, never a deterministic patch. Prefer correctness over reuse. Consider recomputation cost, validation cost, stale-work risk, and how much verified work remains applicable.

\begin{itemize}
\item \texttt{select\_fresh\_solve}: independently solve the complete current task. Prefer this when recomputation is cheap, inherited reasoning is risky, or checking old work costs as much as rebuilding it. Short calculations are often safer to recompute.
\item \texttt{select\_reuse\_and\_patch}: fork the verified base workspace and update only what the complete current task requires. Prefer this when valuable reports, research, evidence, code structure, style, or interfaces would be expensive to recreate. Set \texttt{preserve\_base\_calculations=true} only when verified base calculations are supplied and every supplied calculation remains required and unchanged. It may be true when the task adds new calculations, but must be false if any base calculation is replaced, withdrawn, stale, or no longer required.
\item \texttt{inspect\_base\_workspace}: read one relevant file before deciding. Inspect only when the task comparison and manifest are insufficient. After inspection, select a final strategy. If uncertainty remains, call \texttt{select\_fresh\_solve}.
\end{itemize}

These are considerations, not fixed domain rules. Do not choose reuse merely because a base exists. Do not output Runtime IDs, patches, verification claims, or an answer to the underlying task. A final strategy tool call ends routing.

\textbf{Runtime input:} \texttt{current\_resolved\_task}, \texttt{base\_resolved\_task}, \texttt{change\_hint}, and the base-workspace manifest.

\textbf{Output:} Return exactly one structured tool call per response. Inspection is non-terminal and must be followed by a selection call; either selection call is terminal. Fields marked optional may be omitted, and no additional fields are allowed.

\noindent\texttt{inspect\_base\_workspace}\par
\noindent\texttt{\{"path":"<string>", "offset":<integer $\geq 0$, optional>,}\par
\noindent\texttt{"limit":<integer in [1,12000], optional>,}\par
\noindent\texttt{"reason":"<string>"\}}

\noindent\texttt{select\_fresh\_solve}\par
\noindent\texttt{\{"reason":"<non-empty string>"\}}

\noindent\texttt{select\_reuse\_and\_patch}\par
\noindent\texttt{\{"reason":"<non-empty string>",}\par
\noindent\texttt{"preserve\_base\_calculations":<true or false>\}}
\end{promptbox}

\subsection{Workspace Execution}

The following shared suffix governs how the downstream executor reads and updates the selected workspace. Domain-specific prefixes define the available tools, artifact schema, and completion conditions and are provided separately.

\begin{promptbox}{Shared Workspace Method-Selection Suffix}
Treat \texttt{current\_resolved\_task} as the complete current task specification. Choose the working method yourself before acting. Do not read prior workspace content merely because it exists. First decide whether the task is self-contained and cheap to complete independently. For a small math problem, short question, or small generation task, solve directly from \texttt{current\_resolved\_task} without inspecting prior output. Read prior files only when the task depends on valuable accumulated work such as existing code structure, report content, references, style, or interfaces that would be wasteful or unsafe to recreate. For a localized change to such work, inspect only relevant files and patch the smallest necessary region. For a structural change, use existing files as reference and rewrite only affected artifacts. If existing work already satisfies the task, finish without editing. Do not report this choice as Runtime metadata; express it through your tool actions.

\textbf{Runtime input:} \texttt{current\_resolved\_task}, the selected workspace, its manifest, and any domain-specific execution context.

\textbf{Output:} Return exactly one available structured tool call per response. The shared workspace calls are shown below; a domain may restrict paths or add a task tool such as \texttt{calculate} or \texttt{browsecomp\_search}. No additional fields are allowed.

\noindent\texttt{glob\_workspace}\par
\noindent\texttt{\{"pattern":"<string>"\}}

\noindent\texttt{grep\_workspace}\par
\noindent\texttt{\{"pattern":"<string>",}\par
\noindent\texttt{"paths":<array of strings, optional>\}}

\noindent\texttt{read\_workspace}\par
\noindent\texttt{\{"path":"<string>", "offset":<integer $\geq 0$, optional>,}\par
\noindent\texttt{"limit":<integer in [1,50000], optional>\}}

\noindent\texttt{write\_workspace}\par
\noindent\texttt{\{"path":"<string>", "content":"<string>"\}}

\noindent\texttt{patch\_workspace}\par
\noindent\texttt{\{"path":"<string>",}\par
\noindent\texttt{"replacements":[\{"old\_text":"<string>",}\par
\noindent\texttt{"new\_text":"<string>"\}]\}}

\noindent\texttt{delete\_workspace}\par
\noindent\texttt{\{"path":"<string>"\}}

\noindent\texttt{finish\_work\_memory}\par
\noindent\texttt{\{"final\_response":"<string>"\}}
\end{promptbox}

\subsection{Evaluation Prompt}

The following frozen prompt is used only to evaluate whether the final resolved query expresses the same complete task as the hidden reference. It is not shown to the Git Agent and does not generate the downstream task answer. At runtime, \texttt{\{domain\_specific\_semantic\_check\}} is filled from Table~\ref{tab:query_equivalence_domain_checks}.

\begin{promptbox}{Query-Equivalence Judge Prompt}
Judge whether \texttt{PREDICTION} and \texttt{REFERENCE} express the same complete current task. Account for the requested goal, all active facts, constraints, comparisons, output requirements, withdrawals, and replacements.

Judge semantic effect rather than textual minimality. Equivalent paraphrases are correct. Extra accepted background is allowed only when it is non-conflicting and cannot change the requested goal, answer, scope, or active constraints. Do not apply this allowance to stale or withdrawn clauses, or to facts that could affect the answer or scope.

Missing active clauses, retaining material stale or withdrawn clauses, changing values, or introducing contradictions are not equivalent. Ignore differences in purely decorative step-by-step or boxed-answer wrappers; enforce explicit structured-output fields, precision, and task constraints.

\textbf{Domain-specific semantic check:} \texttt{\{domain\_specific\_semantic\_check\}}

\textbf{Input (JSON):}

\noindent\texttt{\{}\\
\quad \texttt{"prediction": "\{candidate\_resolved\_query\}",}\\
\quad \texttt{"reference": "\{hidden\_reference\}"}\\
\texttt{\}}

\textbf{Output:} Return exactly one \texttt{judge\_query\_equivalence} tool call.

\noindent\texttt{\{}\\
\quad \texttt{"equivalent": <true or false>,}\\
\quad \texttt{"error\_type": "<none | missing | stale | contradictory |
different\_goal>"}\\
\texttt{\}}

Do not solve the task.
\end{promptbox}